\documentclass[english, a4paper, 11pt, DIV=12, numbers=noenddot]{scrartcl}
\usepackage{setspace}
\usepackage[bottom]{footmisc}
\usepackage{blindtext}
\usepackage{amsmath}
\usepackage{amssymb}
\usepackage{bbm}
\usepackage{slashed}
\usepackage{cite}
\usepackage{hyperref}
\usepackage{graphicx}
\usepackage[labelfont=bf]{caption}
\usepackage{subcaption}
\usepackage{braket}
\usepackage{chngcntr}
\usepackage{booktabs}
\usepackage{color, colortbl, xcolor}
\usepackage{multirow}
\numberwithin{equation}{section} 
\counterwithin{figure}{section} 
\counterwithin{table}{section} 
\hypersetup{linktocpage,
    colorlinks,
    linkcolor={red!70!black},
    citecolor={green!50!blue},
    urlcolor={blue!50!black}
}

\newcommand*{\LessSim}{\smallrel\lesssim}

\makeatletter
\newcommand*{\smallrel}[2][.8]{%
  \mathrel{\mathpalette{\smallrel@{#1}}{#2}}%
}
\newcommand*{\smallrel@}[3]{%
  \sbox0{$#2\vcenter{}$}%
  \dimen@=\ht0 %
  \raise\dimen@\hbox{%
    \scalebox{#1}{%
      \raise-\dimen@\hbox{$#2#3\m@th$}%
    }%
  }%
}
\makeatother

\def\EmissT{\,{}/ \hspace{-1.5ex}  E_{T}}

\newcommand{\D}{\mathrm{d}}
\newcommand{\eq}{{\mathrm{eq}}}

\begin{document}
\begin{titlepage}
\begin{flushright}
TTP23-059\\
P3H-23-102\\
TTK-23-35
\end{flushright}
\vskip2cm
\begin{center}
{\LARGE \bfseries 
Flavoured Majorana Dark Matter then and now:\\[2mm]
From freeze-out scenarios to LHC signatures}

\vskip1.0cm
{\large Harun Acaro\u{g}lu$^a$, Monika Blanke$^{a,b}$, Jan Heisig$^{c,d}$, Michael Kr\"amer$^c$, Lena Rathmann$^c$}
\vskip0.5cm
 \textit{$^a$Institut f\"ur Theoretische Teilchenphysik,
  Karlsruhe Institute of Technology, \\
Engesserstra\ss e 7,
  D-76128 Karlsruhe, Germany}
 \vspace{3mm}\\
  \textit{$^b$Institut f\"ur Astroteilchenphysik, Karlsruhe Institute of Technology,\\
  Hermann-von-Helmholtz-Platz 1,
  D-76344 Eggenstein-Leopoldshafen, Germany} \vspace{3mm}\\
    \textit{$^c$ Institute for Theoretical Particle Physics and Cosmology, RWTH Aachen University, Sommerfeldstr. 16, D-52056 Aachen, Germany}\vspace{3mm}\\
  \textit{$^d$Department of Physics, University of Virginia, 382 McCormick Road,
Charlottesville, Virginia 22904-4714, USA}\vspace{3mm}\\


\vskip1cm


\vskip1cm

{\large \bfseries Abstract\\[10pt]} \parbox[t]{.9\textwidth}{
}
\vspace{-3ex}
\begin{abstract}
We study a simplified Dark Matter model in the Dark Minimal Flavour Violation framework. Our model complements the Standard Model with a flavoured Dark Matter Majorana triplet and a coloured scalar mediator that share a Yukawa coupling with the right-handed up-type quarks with the coupling matrix $\lambda$. We extend previous work on this topic by exploring a large range of cosmologically viable parameter space, including the coannihilation region and, in particular, the region of conversion-driven freeze-out, while considering constraints from $D^0-\bar D^0$ mixing as well as constraints from direct and indirect Dark Matter searches. We find various realisations of conversion-driven freeze-out within the model, that open up allowed windows of parameter space towards small mass splittings and very weak Dark Matter couplings. Finally, we probe the model by reinterpreting current LHC searches for missing energy and long-lived particles. We point out gaps in the coverage of current constraints as well as new opportunities to search for the model at the LHC, in particular, the charge asymmetry in single-top production associated with jets and missing energy.
\end{abstract}
\end{center}
\end{titlepage}

\tableofcontents 

\section{Introduction}
\label{sec:intro}

As of today, one of the strongest indications for the presence of New Physics (NP) beyond the Standard Model (SM) is the evidence for the existence of Dark Matter (DM) in our Universe. While the DM mass scale is \emph{a priori} a free parameter, a popular assumption for its particle nature is the so-called WIMP -- a weakly interacting massive particle with a mass around the electroweak scale. Indeed, this option is well-motivated both theoretically and phenomenologically. On the one hand, the presence of new weakly coupled particles around the TeV scale is typically required in NP models addressing the hierarchy problem, providing the exciting opportunity to link the issue of naturalness to the DM puzzle. On the other hand, the WIMP paradigm predicts the DM relic abundance in the right ballpark for weakly coupled DM candidates in this mass range. Furthermore, WIMP DM can be tested both by current DM direct detection experiments and by particle physics experiments at the Large Hadron Collider (LHC)\@.
Indeed, a variety of such DM searches have been performed over the past years, but so far failed to provide any direct evidence for DM\@. These null results, in turn, place stringent limits on the parameter space of DM and put simple WIMP models under severe pressure.

An appealing solution to this tension is provided by the introduction of a non-trivial flavour structure into the dark sector. Flavoured DM (FDM) models assume the presence of three matter generations in the dark sector together with their flavour-violating coupling to the SM matter~\cite{Kile:2011mn,Kamenik:2011nb,Batell:2011tc,Agrawal:2011ze,Batell:2013zwa,Kile:2013ola,Kile:2014jea, Lopez-Honorez:2013wla}.  While early FDM models have been restricted to the Minimal Flavour Violation  hypothesis~\cite{Buras:2000dm,DAmbrosio:2002vsn,Chivukula:1987py,Hall:1990ac}, over the past decade a series of studies of FDM models beyond Minimal Flavour Violation have been carried out in both the quark~\cite{Agrawal:2014aoa,Blanke:2017tnb,Blanke:2017fum,Jubb:2017rhm,Acaroglu:2021qae} and lepton~\cite{Chen:2015jkt,Acaroglu:2022hrm,Acaroglu:2022boc,Acaroglu:2023cza,Asadi:2023csb} sectors. A convenient choice is the assumption of Dark Minimal Flavour Violation (DMFV)~\cite{Agrawal:2014aoa} which allows for a single new source of flavour violation beyond the SM Yukawa couplings. While capturing the phenomenological effects of new flavour-violating interactions, DMFV keeps the number of new parameters manageable.

Previous studies of DMFV models have provided a broad picture of their phenomenology. Relevant constraints arise from the observed relic abundance in the Universe, flavour-violating observables, DM searches in direct and indirect detection experiments and at colliders\@. In terms of the LHC discovery prospects, quark-flavoured DM models are much more promising than their lepton-flavoured counterparts due to the QCD charge of the mediating particles implying sizeable production cross sections. In addition to the flavour-conserving di-quark and missing transverse energy ($\EmissT$) final states that are constrained by standard LHC searches for supersymmetric squarks, the flavour structure of the dark sector also induces flavour-violating final states. The latter are most easily accessible in models that couple the dark sector to the up-type quarks of the SM, since top quarks are experimentally distinct from light quark jets. The resulting single-top final states have been shown to significantly extend the reach of LHC searches~\cite{Blanke:2020bsf}.

Apart from the type of SM fermions to which the dark sector couples, FDM models are also characterised by the particle nature of the DM flavour multiplet and the mediator particle. While in most previous analyses, these have been assumed to be Dirac fermions and a scalar boson, respectively, in Ref.~\cite{Acaroglu:2021qae} the case of Majorana DM coupling to right-handed up-type quarks via a scalar mediator has been investigated. The phenomenological implications can be directly compared with those of the analogous model with Dirac DM~\cite{Blanke:2017tnb}, exhibiting crucial differences in all aspects of the model's phenomenology. On the one hand, mediator pair-production at the LHC receives a significant enhancement by Majorana-specific same-sign $t$-channel contributions. The latter also gives rise to Majorana-specific signatures which we explore further in Sec.~\ref{sec:LHC}, 
including the proposal of a novel charge asymmetry that appears to be a promising tool to test the nature of DM\@. On the other hand, signals in direct and indirect detection experiments get suppressed in the Majorana case. At the same time, the flavour phenomenology gets altered due to additional contributions to the constraining neutral meson mixing process, thus opening up parameter space for new phenomena in rare and CP-violating processes. Last but not least, also the DM freeze-out is qualitatively affected by the Majorana nature of DM, leading to a $p$-wave suppression of DM annihilation into light final states.
In the present paper, we revisit the  Majorana FDM model of Ref.~\cite{Acaroglu:2021qae} and explore in greater depth several aspects of its phenomenology. 

Previous analyses of FDM models are typically restricted to two limiting benchmark cases for the DM freeze-out in the early Universe, assuming either a significant mass splitting in the dark sector, such that only the lightest, stable flavour contributes to the thermal annihilation cross section (``single flavour freeze-out''), or a negligible mass splitting between the dark flavours, such that they all yield a relevant contribution (``quasi-degenerate freeze-out''). In the present work, we instead consider the generic case which allows us to cover significant additional parameter space with qualitatively different phenomenology. On the one hand, we systematically study the effect of intermediate mass splitting between the various NP states, which leads to sizeable coannihilation effects in the process of DM freeze-out. On the other hand, we explore a wide range of couplings between the NP particles that go beyond the validity of the assumptions commonly made within the WIMP paradigm. In particular, we examine a variety of realisations of conversion-driven freeze-out~\cite{Garny:2017rxs,DAgnolo:2017dbv} within the model. In this freeze-out scenario, the coupling of the DM state can be orders of magnitude weaker than that of a WIMP, consistent of course with the null-results in canonical WIMP searches. At the same time, the heavier states of the DM multiplet and/or the scalar mediator can drive efficient annihilation in the early Universe due to sizeable couplings to the SM\@. This opens up several interesting variations of the model that lead to cosmologically viable freeze-out scenarios that have not yet been explored. 

The computation of the relic density is more involved than in the case of a WIMP. The chemical equilibrium between the NP particles cannot be assumed in conversion-driven freeze-out, since the chemical decoupling of DM is initiated by semi-efficient conversion processes between DM and the heavier NP states. In fact, the chemical decoupling within the NP sector generally interferes with the decoupling of the NP sector from the SM bath~\cite{Garny:2017rxs}. Hence, to compute the relic density, a coupled set of (four) Boltzmann equations has to be solved.

The remainder of this paper is structured as follows. In Sec.~\ref{sec:model}, we introduce the model under consideration. In Sec.~\ref{sec:relic}, we detail the DM relic density computation and discuss   possible freeze-out scenarios. In Sec.~\ref{sec:scans}, we perform several phenomenological scans in the identified scenarios. Finally, collider signatures and constraints are studied in Sec.~\ref{sec:LHC} before concluding in Sec.~\ref{sec:Conclude}. Appendix~\ref{app:XS} provides analytical expressions for the annihilation and conversion rates used in the relic density computation while in App.~\ref{qpp:abund} we show exemplary solutions of the Boltzmann equations for the evolution of the particle abundances.

\section{Particle physics model}
\label{sec:model}

The simplified Majorana FDM model that we consider in the present analysis was originally introduced in Ref.~\cite{Acaroglu:2021qae}. DM is represented by a Majorana fermion $\chi$ which is a singlet under the SM gauge group but transforms as a flavour triplet under a new approximate global flavour symmetry $O(3)_\chi$ in the dark sector. Its coupling to the right-handed up-type quarks of the SM is mediated by a scalar field $\phi$ which carries the same gauge quantum numbers as the up-type quarks, \emph{i.e.}~$\phi \sim (\mathbf{3},\mathbf{1},2/3) $. To ensure DM stability, the new fields are further charged under a discrete $\mathbb{Z}_2$ symmetry. The Lagrangian is given by
\begin{eqnarray}
    \mathcal{L}&=& \mathcal{L}_\text{SM} + \frac{1}{2}\left(i\bar\chi\slashed\partial\chi - M_\chi \bar\chi\chi\right)
    -\left(\lambda_{ij}\bar u_{Ri} \chi_j\phi + \text{h.c.}\right) \nonumber\\
    &&+(D_\mu \phi)^\dagger (D^\mu\phi) -m_\phi^2\phi^\dagger\phi  +\lambda_{H\phi}\phi^\dagger\phi  H^\dagger H +\lambda_{\phi\phi}\left(\phi^\dagger\phi \right)^2\,,
\end{eqnarray}
where $\chi$ is a four-component Majorana spinor.
The model is assumed to obey the Dark Minimal Flavour Violation (DMFV) hypothesis, which requires the flavour-violating coupling matrix $\lambda$ to be the only new source violating the flavour symmetry of the model  
\begin{equation}
    \mathcal{G}_\text{flavour} = U(3)_Q \times U(3)_u \times U(3)_d \times O(3)_\chi\,.
\end{equation}
The DM mass matrix $M_\chi$ can thus be written as
\begin{equation}
\label{eq:masscorr}
M_\chi = m_\chi \left[\mathbbm{1} + \eta \,\text{Re}(\lambda^\dagger\lambda) + \mathcal{O}(\lambda^4) \right]\,,
\end{equation}
in terms of the DMFV spurion expansion. Note that the symmetry of the Majorana mass matrix $M_\chi$ implies the presence of the real part $\text{Re}(\lambda^\dagger\lambda)$ in the spurion expansion, instead of $\lambda^\dagger\lambda$ present for flavoured Dirac DM~\cite{Agrawal:2014aoa,Blanke:2017tnb}.

The flavour-violating coupling matrix $\lambda$ is a generic complex $3\times 3$ matrix. However, the flavour symmetry $O(3)_\chi$ renders three of its parameters unphysical, which can hence be removed by appropriate field redefinitions. The remaining {15} physical parameters of 
\begin{equation}
    \lambda = U D\, O\, d\,,
\end{equation} are then
\begin{itemize}
\item three mixing angles $\theta_{ij}$ and three complex CP phases $\delta_{ij}$ of the unitary matrix $U$,
\item three real non-negative coupling parameters $D_i$, $i=1,2,3$ which form the diagonal matrix $D$,
\item three mixing angles $\phi_{ij}$ of the orthogonal matrix $O$, and
\item three complex phases $\gamma_i$ of the diagonal matrix $d$.
\end{itemize}
The details of the parametrisation can be found in Ref.~\cite{Acaroglu:2021qae}.
Note that the matrices $O$ and $d$ are present only in the Majorana case, due to the smaller flavour symmetry compared to the Dirac case studied in Ref.~\cite{Blanke:2017tnb}. 

For the same reason, in the Majorana case the DM mass matrix from Eq.~\eqref{eq:masscorr} is not diagonal per se, \emph{i.e.}~it has to be diagonalised through an Autonne-Takagi factorisation~\cite{takagi,autonne}, {$M_\chi = W^T M_\chi^D W$, where $M_\chi^D = \text{diag}(m_{\chi_1},m_{\chi_2},m_{\chi_3})$ with $m_{\chi_1} > m_{\chi_2} > m_{\chi_3}$ and $W$ is an orthogonal matrix. Thus the third dark generation is the lightest state and forms the DM of the Universe.} This procedure transforms the coupling of the DM triplet $\chi$ to the up-type quarks $u_R$ into
\begin{equation}
\label{eq:lambdatilde}
    \tilde\lambda = \lambda W^T\,.
\end{equation}

The quartic couplings $\lambda_{H\phi}$ and $\lambda_{\phi\phi}$ in the extended scalar sector are not relevant for the remainder of our analysis and we set them to zero in what follows.

\section{Dark matter relic density }
\label{sec:relic}

In this section we discuss the computation of the DM relic density, extending the standard thermal WIMP scenario by coannihilation and conversion effects present in our model. We then introduce various scenarios capturing the different possible freeze-out dynamics.

\subsection{Boltzmann equations}
\label{sec:BME}
To compute the relic density of DM, we solve the Boltzmann equations for number densities $n_i$ of species $i$ in the early Universe. Since the model contains four mass eigenstates in the NP sector, ${\chi_{1,2,3}}$ and $\phi$, this is generally a set of four coupled differential equations. Considering the comoving number density $Y = n/s$ (where $s$ is the entropy density) and using $x = m_{\chi_3}/T$ to parameterise its evolution in time, the Boltzmann equations read:
\begin{align} 
    \begin{split}
    \frac{\D Y_{\chi_i}}{\D x} = \, & \frac{1}{3H} \frac{\D s}{\D x} \left[ \,\sum_{j = 1}^3 \langle \sigma v \rangle _{\chi_i \chi_j \rightarrow q q'} \left(Y_{\chi_i} Y_{\chi_j} - Y_{\chi_i}^\eq Y_{\chi_j}^\eq\right)   + \langle \sigma v \rangle_{\phi \chi_i \rightarrow q g} \left(Y_{\chi_i} Y_{\phi} - Y_{\chi_i}^\eq Y_{\phi}^\eq\right)\right. \\
    &\hspace{8.2ex} +  \frac{1}{s} \sum_{j \neq i} \left(\Gamma_{\chi_i q \rightarrow \chi_j q'} + \widetilde{\Gamma}_{\chi_i \rightarrow \chi_j q q'}\right) \left( Y_{\chi_i} - Y_{\chi_j}\frac{Y_{\chi_i}^\eq}{Y_{\chi_j}^\eq} \right) \\
    & \hspace{8.2ex}  - \left. \frac{1}{s} \left(\widetilde{\Gamma}_{\phi \rightarrow \chi_i q} + \Gamma_{\phi g \rightarrow \chi_i q} + \Gamma_{\phi q \rightarrow \chi_i g} \right)\left(Y_{\phi} - Y_{\chi_i}\frac{Y_{\phi}^{\mathrm eq}}{Y_{\chi_i}^{\mathrm eq}} \right)\, \right]\,,
    \end{split}
    \label{eq:BME_full_chi}
    \\
    \begin{split}
    \frac{\D Y_{\phi}}{\D x} = \, & \frac{1}{3H} \frac{\D s}{\D x} \left[\, \frac{1}{2} \langle \sigma v \rangle_{\phi \phi \rightarrow gg, q q'}  \left(Y_\phi^2 - {Y_\phi^\eq}^2\right) 
    + \sum_{i} \langle \sigma v \rangle_{\phi \chi_i \rightarrow q g} \left(Y_{\phi} Y_{\chi_i} -  Y_{\phi}^\eq Y_{\chi_i}^\eq\right) \right. \\
    &  \hspace{8.2ex} + \frac{1}{s}  \sum_{i=1}^3 \left. \left(\widetilde{\Gamma}_{\phi \rightarrow \chi_i q} + \Gamma_{\phi g \rightarrow \chi_i q} + \Gamma_{\phi q \rightarrow \chi_i g}\right) \left(Y_{\phi} - Y_{\chi_i}\frac{Y_{\phi}^\eq}{Y_{\chi_i}^\eq} \right) \,\right]\,.
    \end{split}
    \label{eq:BME_full_phi}
\end{align}
The first line of the right-hand side of each equation describes the annihilation processes, where $\langle \sigma v \rangle_{ij\to kl}$ denotes the thermally averaged cross section times M\o ller velocity for the process $ij\to kl$ (see App.~\ref{app:XS}). The following lines describe conversion processes via scattering with rates $\Gamma_{ij\to kl} = \langle \sigma v \rangle_{ij\to kl}\, n_{j}^\eq$ and via decay, where $\widetilde{\Gamma}_{i\to \{k\}} = \Gamma_{i\to \{k\}} K_1(m_i/T)/K_2(m_i/T)$ is the thermally averaged decay rate of the species $i$ decaying into particles $\{k\}$. Here $K_n$ is the modified Bessel function of the second type of order $n$. The equilibrium number density of species $i$ is given by
\begin{equation}
    Y_i^\eq =  \frac{n_i^\eq}{s} =\frac{T m_i^2 g_i}{2 \pi^2 s}   K_2\left( \frac{m_i}{T} \right)\,,
\end{equation}
where $g_i$ are its internal degrees of freedom. In Eqs.~\eqref{eq:BME_full_chi} and~\eqref{eq:BME_full_phi}, $Y_\phi$ is the sum of the abundances of the mediator and its antiparticle and $\langle \sigma v \rangle_{\phi \phi \rightarrow gg, q q'}$ is a shorthand notation for the sum $\langle \sigma v \rangle_{\phi \phi^\dagger \rightarrow gg}+\langle \sigma v \rangle_{\phi \phi^\dagger \rightarrow qq'^\dagger}+\langle \sigma v \rangle_{\phi \phi \rightarrow qq'}$ where for each process we have summed over the final state quark flavours. Analytical expressions for the cross sections and decay rates involved are given in App.~\ref{app:XS}.

Assuming that the particles of the NP sector, $\chi_i,\phi$, are in chemical equilibrium with each other,
\begin{equation}
\frac{Y_{\chi_i}}{Y_{\chi_i}^\eq} = \frac{Y_{\chi_j}}{Y_{\chi_j}^\eq} =
\frac{Y_{\phi}}{Y_{\phi}^\eq}\,,
\end{equation}
we can apply the well-known coannihilation approximation, which reduces the above set of equations to a single Boltzmann equation for the sum $Y=\sum_i Y_{\chi_i} + Y_\phi$. The corresponding collision operator then contains an effective annihilation cross section only~\cite{Edsjo:1997bg}. 

This approach is viable if the conversion rates involved are large compared to the Hubble expansion rate, $\Gamma_\text{conv}\gg H$, throughout the freeze-out process. However, since we are interested in exploring the general case including the regime of conversion-driven freeze-out, we explicitly consider the coupled system Eqs.~\eqref{eq:BME_full_chi} and~\eqref{eq:BME_full_phi}. Nevertheless, whenever justified, to improve numerical stability, we reduce this system to a suitable subset of equations by considering the sum over a subset of states among which we can assume chemical equilibrium due to highly efficient conversions. In this case, the Boltzmann equations for this subsector contain an effective annihilation cross section and an effective conversion rate for each NP state that does not belong to this subsector. A similar approach was considered in Ref.~\cite{Heeck:2022rep}.

\subsection{Freeze-out scenarios}
\label{sec:freezeout_scenarios}

The rich structure of the considered model allows for a variety of cosmologically viable scenarios, \emph{i.e.}~scenarios that explain the measured relic density, $\Omega_{\mathrm{DM}} h^2 = 0.12$~\cite{Planck:2018vyg}.  In this subsection we present a qualitative discussion of the  different possibilities. 

\paragraph{Canonical freeze-out.} For sizeable couplings of all $\chi_i$, the DM relic density is generated via canonical freeze-out. For small mass splittings between DM and (one or some of) $\chi_{1},\chi_{2},\phi$, coannihilation effects can be significant. In particular, these effects allow one to consider smaller DM couplings while still ensuring sufficient dilution of DM, provided that the annihilation cross section of the coannihilator(s) involved is sufficiently large. We refer to this case as the \emph{canonical freeze-out} scenario. This scenario is divided into further sub-scenarios in Sec.~\ref{sec:scan:canonic}.

\paragraph{Conversion-driven freeze-out.} The DM coupling can be so small that its contribution to the DM dilution is completely negligible compared to the pair annihilation of the coannihilators. In this case, the annihilation cross section of the coannihilators dominates the total dilution of the NP particles, and the relic density appears to become independent of the coupling of the DM state. However, the  latter is only true as long as conversions between DM and coannihilators are sufficiently efficient to maintain chemical equilibrium in the NP sector. For very weak DM couplings, the conversion rates become inefficient and initiate DM freeze out. Within this \emph{conversion-driven freeze-out}~\cite{Garny:2017rxs} scenario, a dependence of the relic on the DM coupling is reintroduced. 

In general, both the heavier states $\chi_i$ or $\phi$ can act as coannihilators. In the case of the former, the realisation of a conversion-driven freeze-out requires a certain hierarchy in the Yukawa couplings of $\chi_i$. While the coupling of a coannihilating state $\chi_i$ needs to be large, that of DM -- which governs the conversion rates -- needs to be very weak, typically of the order of $10^{-6}$~\cite{Garny:2017rxs} (or somewhat larger if DM couples predominantly to the top quark~\cite{Garny:2018icg}).  Interestingly, the pattern of mass splitting within the DM multiplet is related to the coupling matrix in the considered model, see Eq.~\eqref{eq:masscorr}. Furthermore, due to the helicity suppressed $\chi_i$ pair-annihilation process into light quarks, its cross section is significantly larger for a coannihilator $\chi_i$ that couples predominantly to the top. Note that in general more than one $\chi_i$ state can be very weakly coupled. Therefore a variety of phenomenologically different realisations occur.

In the following, we introduce a naming scheme in which we specify the (dominant) coannihilator as well as the number of very weakly coupled states and their flavour structure. C$_{\!X} n_f$ denotes a scenario with a coannihilator $X$ (either a heavy $\chi$ state or $\phi$) and $n$ very weakly coupled states $\chi_i$ with dominant coupling to the quark flavour $f$. In Tab.~\ref{tab:nscheme} we illustrate the naming scheme by listing the different scenarios. Note that the list is not exhaustive, but includes all the limiting cases discussed here.

\begin{table}
    \centering
    \caption{Naming scheme of scenarios within the conversion-driven freeze-out regime considered in this paper. The upper part of the table collects scenarios where the freeze-out is driven by $\chi_i$--$\chi_j$ conversions, while the lower part shows scenarios where the freeze-out is driven by $\chi_i$--$\phi$ conversions. Note that the `$\ll$' symbol used in the mass hierarchy display is intended to indicate sufficient splitting to suppress a significant contribution from coannihilation effects. For the three scenarios marked with an asterisk, we present a detailed analysis.}
    \vspace{2mm}
    \begin{tabular}{ccccc}
    \toprule
    &\textbf{name} & \textbf{mass hierarchy} & \textbf{very weakly coupled} & \textbf{dominant DM coupling} \\ 
    \bottomrule 
    \toprule
      $\star\!\!$  & C$_\chi 1_u$ & $m_{\chi_3} < m_{\chi_{1,2}}\ll m_\phi  $ & $\chi_3$ & $\tilde\lambda_{u3}$ \\
       & C$_\chi 1_t$ & $m_{\chi_3} < m_{\chi_{1,2}}\ll m_\phi  $ & {$\chi_3$} & $\tilde\lambda_{t3}$ \\
       &  C$_\chi 2_u$ & $m_{\chi_{2,3}} < m_{\chi_{1}} \ll m_\phi$ & $\chi_2,\chi_3$ & $\tilde\lambda_{u3}$\\
        \midrule
       $\star\!\!$ &C$_\phi 1_u$ & $m_{\chi_3} < m_\phi \ll m_{\chi_{1,2}}$ & $\chi_3$ & $\tilde\lambda_{u3}$\\
        &C$_\phi 1_t$ & $m_{\chi_3} < m_\phi \ll m_{\chi_{1,2}}$ & $\chi_3$ & $\tilde\lambda_{t3}$\\
       $\star\!\!$ &C$_\phi 2_u$ & $m_{\chi_{2,3}} < m_\phi \ll m_{\chi_{1}}$ & $\chi_2,\chi_3$ & $\tilde\lambda_{u3}$ \\
    \bottomrule
    \end{tabular}
    \label{tab:nscheme}
\end{table}

\section{Phenomenological scans}
\label{sec:scans}

\subsection{General scan setup}

In this section, we perform a global analysis in which we determine the valid parameter space of our model using constraints from flavour observables, direct and indirect detection, as well as the relic density. Current constraints from collider searches and future prospects are discussed in Sec.~\ref{sec:LHC}. With respect to the relic density calculation, we distinguish between a canonical WIMP freeze-out and a conversion-driven freeze-out scenario, described in Sec.~\ref{sec:scan:canonic} and Sec.~\ref{sec:scan:conversion}, respectively.  In both scenarios, we require the predicted relic density to match the experimentally measured value, $\Omega h^2 = 0.120 \pm 0.001$~\cite{Planck:2018vyg}, within an assumed theoretical uncertainty of $10\%$. Our analysis of the constraints from the flavour and (in)direct detection experiments can be summarised as follows:

\paragraph{Flavour constraints.}
The flavour structure of the coupling matrix $\tilde\lambda$ is subject to strong constraints from flavour observables. Since in our model DM is coupled to up-type quarks, the relevant process in this case consists of $D^0$--$\bar D^0$ mixing with $\Delta C = 2$. Most importantly, the Majorana nature of DM gives rise to an additional contribution to this process, where in contrast to the usual box diagram the fermion lines of the dark particles in the loop are crossed leading to a relative sign between both contributions~\cite{Acaroglu:2021qae}. Thus, this additional contribution can lead to destructive interference, opening up important parameter space. However, this feature comes at the cost of tighter constraints outside the interference region: the crossed fermion line diagram is proportional to the diagonal elements of $\tilde\lambda$, and thus it constrains the couplings $D_1$ and $D_2$ even in the flavour-conserving case with $\theta_{ij}=\phi_{ij}=0$. We refer the reader to Ref.~\cite{Acaroglu:2021qae} for more details. To constrain the $D^0$ meson system we use the numerical values given in Refs.~\cite{lattice1,lattice2,mesonscale,pdg,mixinglimits}.

\paragraph{Direct detection constraints.} The absence of a signal in direct detection experiments constrains the parameter space of our model. For Dirac DM, the strongest constraints typically arise from spin-independent DM-nucleon scattering, since this process is coherently enhanced by DM scattering from all nucleons in the nucleus. In our model, however, the leading dimension-six operators $\bar\chi\chi \bar q q$ and $\left(\bar\chi\gamma^\mu\chi\right)\left(\bar q\gamma_\mu q\right)$ vanish due to the chiral coupling structure and the DM being a Majorana fermion. A proper analysis of the direct detection constraints therefore requires not only the consideration of spin-dependent DM-nucleon scattering, but also the inclusion of loop-induced DM-gluon scattering contributions to the spin-independent scattering cross section. A full description of how both the spin-dependent and spin-independent scattering cross sections are calculated can be found in Ref.~\cite{Acaroglu:2021qae}. Here we use updated limits obtained from Ref.~\cite{LZ:2022lsv}.

\paragraph{Indirect detection constraints.}
In the DM mass range between 200\,GeV and 1\,TeV, cosmic-ray antiprotons provide a particularly strong constraint on the annihilation cross section of DM in our galaxy. Note that the $s$-wave cross section of the processes $\chi_3 \chi_3 \to q \bar q$ is helicity suppressed. Accordingly, only annihilation in $t\bar t$ proceeds at a significant rate in our galaxy, since the $p$-wave contribution to $\sigma v$ is proportional to $v^2\sim 10^{-6}$. Therefore, we compute the cross section $\chi_3 \chi_3 \to t \bar t$ and confront it with the corresponding limits taken from Ref.~\cite{Cuoco:2017iax}.

\bigskip
\medskip

In the global analysis we scan over all 18 free parameters of our model, \emph{i.e.}~the 15 parameters contained in the coupling matrix $\lambda$ and described in Sec.~\ref{sec:model}, the parameter $\eta$ from Eq.~\eqref{eq:masscorr} which governs the DM mass corrections, as well as the mass parameters $m_\chi$ and $m_\phi$. The scan ranges depend on the different freeze-out scenarios considered below, and are therefore defined and motivated separately for each scenario. To determine the allowed parameter space of our model, for each randomly generated point we first compute the DM mass matrix $M_\chi$ via Eq.~\eqref{eq:masscorr}, diagonalise it and then determine the coupling matrix $\tilde\lambda$ as defined in Eq.~\eqref{eq:lambdatilde}. Using $\tilde\lambda$ we then calculate all the relevant observables discussed above and compare them with their experimental measurements or upper bounds to determine if the point is allowed or not.  Note that among the 18 free model parameters, the most relevant ones are $m_\chi$, $m_\phi$, $D_{1,2,3}$, and $\eta$ while the complex phases and mixing angles entering $\lambda$ have a minor effect on the phenomenology considered here, as long as they lie within the predefined ranges allowed for a given coupling scenario.

\subsection{Canonical freeze-out scenario}
\label{sec:scan:canonic}
In this scenario, we consider large couplings and mixing angles of the $\chi_i$ states, which make the conversion rates between all $\mathbbm{Z}_2$-odd particles efficient, so that chemical equilibrium between them is maintained during the freeze-out. Accordingly, the DM genesis proceeds via the canonical WIMP freeze-out in the early Universe. Based on the mass hierarchy of $\mathbbm{Z}_2$-odd particles, this scenario can be further divided into sub-scenarios, as already done in Ref.~\cite{Acaroglu:2021qae}. There, only two discrete freeze-out scenarios were considered, which were defined as follows:
\begin{itemize}
    \item The \textit{Quasi-Degenerate Freeze-Out} (QDF) scenario, where the mass splittings
    \begin{equation}
\Delta m_{i3} = \frac{m_{\chi_i}}{m_{\chi_3}} - 1\,,
    \end{equation}
    between the lightest and the heavier dark flavours $i\in\{1,2\}$ are assumed to be smaller than $1\%$. For such small mass splittings, the decay of the heavy flavours into lighter states is kinematically suppressed, such that all dark flavours are present at and contribute to the thermal freeze-out of DM\@. 
    \item The \textit{Single Flavour Freeze-Out} (SFF) scenario, where the mass splittings $\Delta m_{i3}$ are assumed to be larger than $10\%$. In this case, flavour-changing scatterings of the form $\chi_i u_k \to \chi_j u_l$ still maintain a relative equilibrium between all dark flavours, however the number densities of the heavy states $\chi_1$ and $\chi_2$ are severely Boltzmann suppressed. Thus, to an excellent approximation, the freeze-out is driven only by annihilations of the lightest state $\chi_3$.
\end{itemize}
In Ref.~\cite{Acaroglu:2021qae}, these definitions are further complemented by the global assumptions that the DM mass corrections are negative, \emph{i.e.}~$\eta <0$ and that the mediator $\phi$ is decoupled, \emph{i.e.}~$\Delta m_{\phi 1} = m_{\phi}/m_{\chi_1} - 1 >10\%$. In our current analysis, however, we perform a more generic scan in which we do not constrain the mass hierarchy of the dark sector in any way other than the chosen parameter ranges, \emph{i.e.}~we neither constrain the sign of $\eta$ to be negative, nor reject points based on $\Delta m_{i3}$ or $\Delta m_{\phi 1}$. We call this case \textit{Canonical Freeze-Out}. In order to allow a direct comparison of our results with those of Ref.~\cite{Acaroglu:2021qae}, we define a third sub-scenario called the \textit{Generic Canonical Freeze-Out} (GCF) scenario, in which we exclude from the canonical freeze-out those parameter points that fall into the QDF and SFF scenarios. The canonical freeze-out is thus split into three sub-scenarios: QDF, SFF and GCF, for which we present the results separately. 

Due to the efficient conversions, the Boltzmann equations from Eqs.~\eqref{eq:BME_full_chi} and \eqref{eq:BME_full_phi} can be reduced to a single one, allowing the use of standard tools to compute the DM relic density. In practice, we use \textsc{micrOMEGAs}~\cite{Belanger:2018ccd} for this calculation. Note that this is another difference from the analysis performed in Ref.~\cite{Acaroglu:2021qae}, where the partial wave expansion coefficients $a$ and $b$ of the DM annihilation rate $\langle \sigma v\rangle = a+b\,\langle v^2\rangle$ were calculated and compared with literature values. 

In each scenario we randomly sample points according to the scan ranges and priors listed in Tab.~\ref{tab:ranges:canonic}. We only keep points that satisfy all constraints, applying the relic density constraint in the last step as it requires the most time consuming computation. We accumulate $10^4$ points per sub-scenario. The fraction of allowed points out of the total number of points sampled is of the order of $10^{-4}$.

\begin{table}[htb]
	\centering
    \caption{Scan ranges for the canonical freeze-out scenario. The first column displays the parameter, the second column its scan range and the third the respective prior, \emph{i.e.}~it indicates whether the scan range is sampled linearly (lin) or logarithmically (log). The upper part of the table shows the parameters 
    contained in the coupling matrix $\tilde\lambda$, while the lower part shows the mass parameters.}
    \vspace{2mm}
	\begin{tabular}{ccc} 
    \toprule
    \textbf{parameter}& \textbf{range}& \textbf{prior} \\ 
    \bottomrule 
    \toprule
	$D_i$ & $[0,2]$ & lin \\ 
    $\theta_{ij}$& $[0,\pi/4]$& lin\\
    $\phi_{ij}$& $[0,\pi/4]$& lin\\
    $\delta_{ij}$& $[0,2\pi)$& lin\\
    $\gamma_i$& $[0,2\pi)$& lin\\
    \midrule
    $\eta$& $[-1,-0.01]\cup [0.01, 1]$& log\\ 
    $m_\chi\,[\mathrm{GeV}]$& $[100,2000]$& lin\\
    $m_\phi\,[\mathrm{GeV}]$& $[m_{\chi_3},m_{\chi_3}+2000]$& lin\\
    \bottomrule
	\end{tabular}
	\label{tab:ranges:canonic}
\end{table}

\begin{figure}
	\centering
	\begin{subfigure}[t]{0.49\textwidth}
        \centering
		\includegraphics[width=\textwidth]{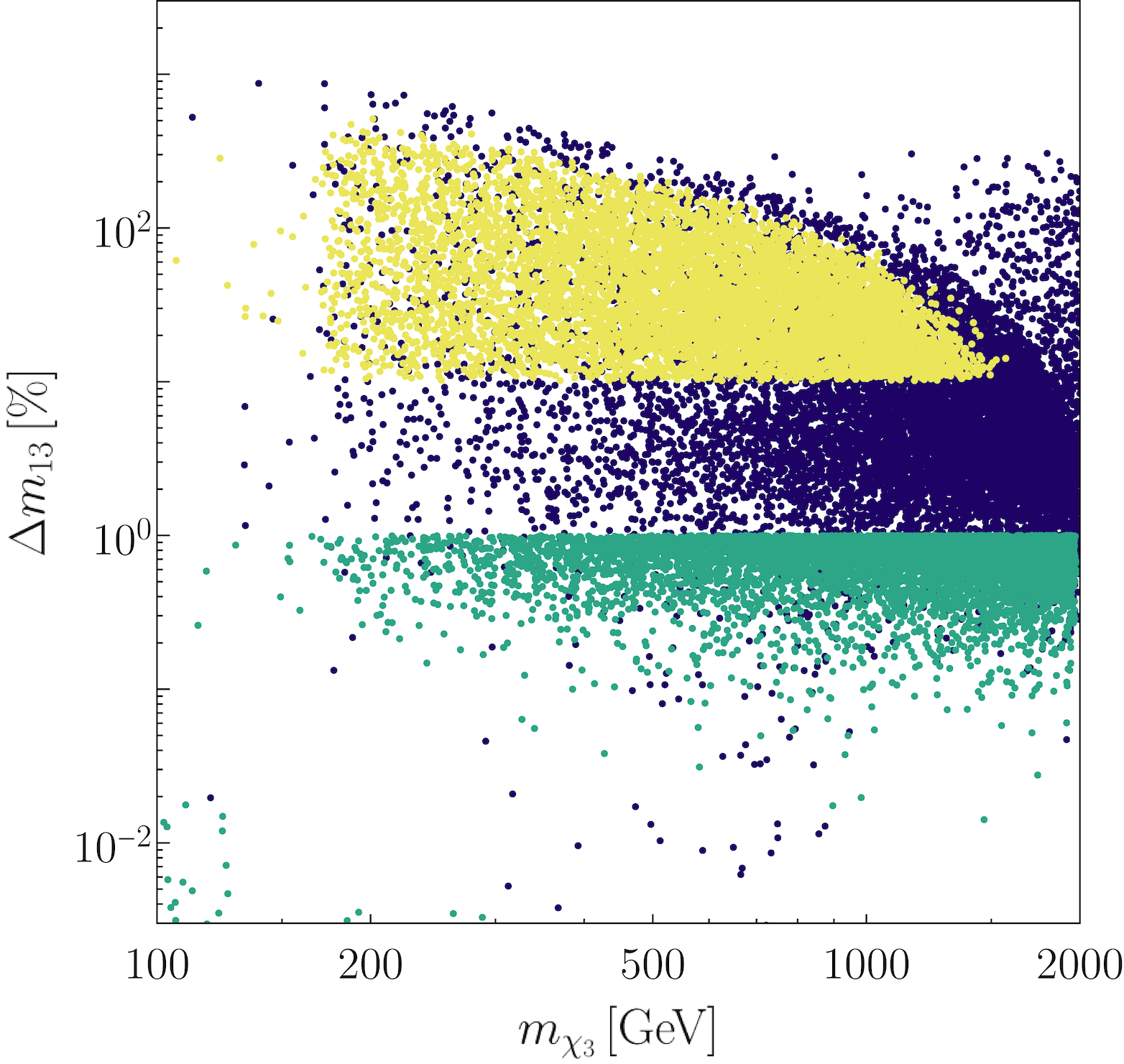}
        \label{fig:canonic:mphimchi3}
	\end{subfigure}
	\hfill
        \begin{subfigure}[t]{0.49\textwidth}
        \centering
		\includegraphics[width=\textwidth]{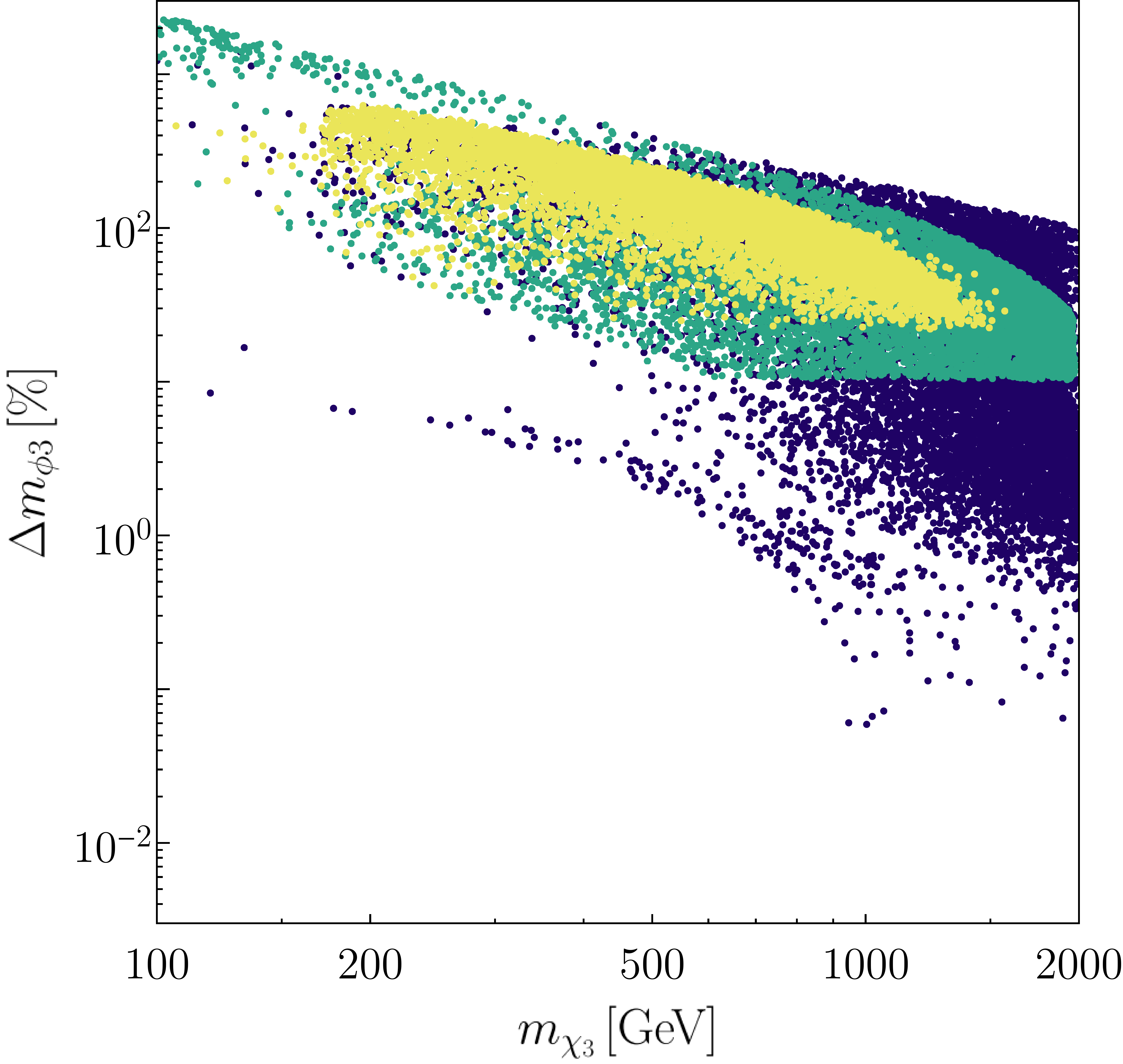}
        \label{fig:canonic:mchi3Dm13}
	\end{subfigure}
     \begin{subfigure}[t]{0.49\textwidth}
        \centering
		\includegraphics[width=\textwidth]{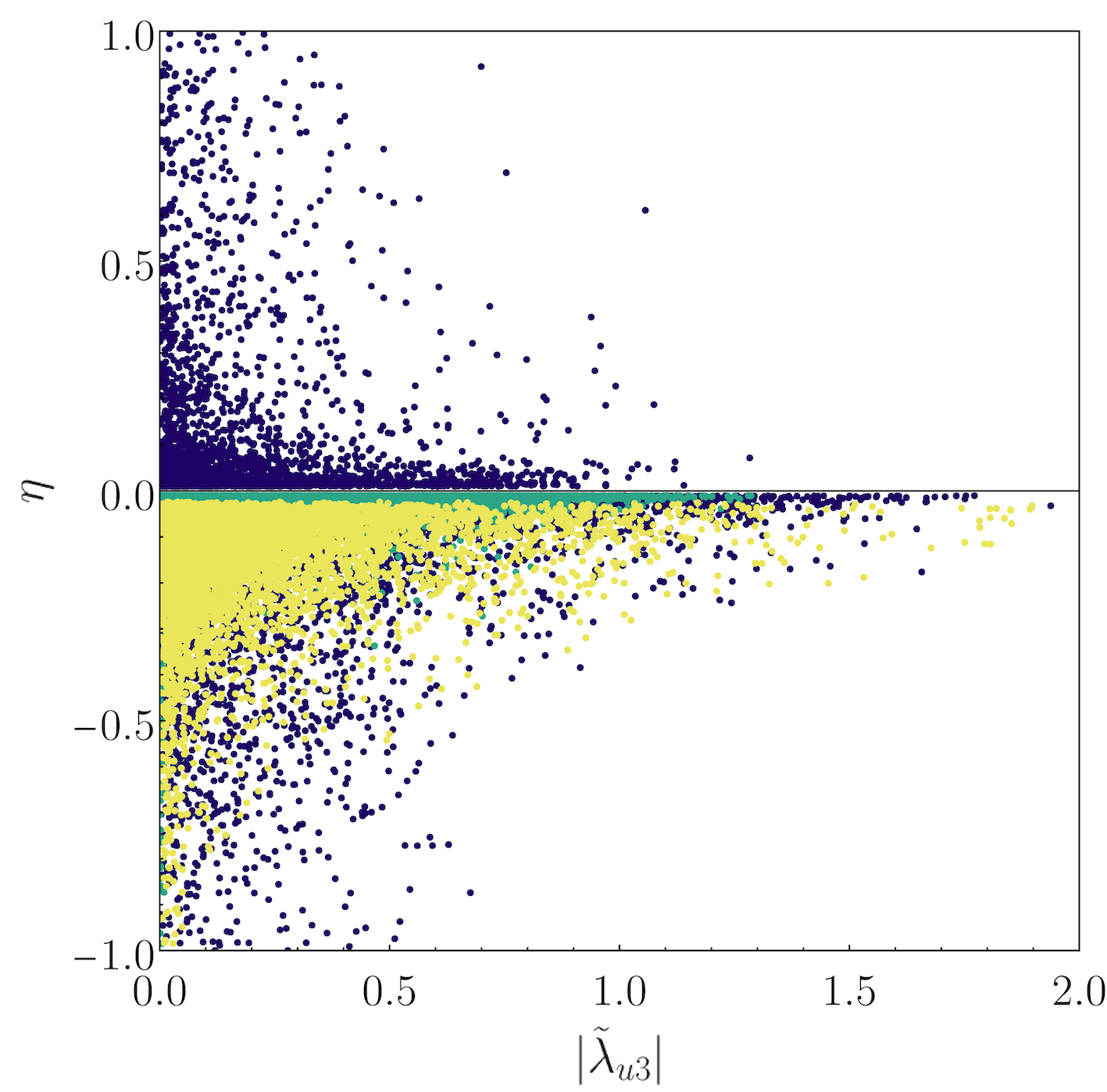}
        
        \label{fig:canonic:lu3eta}
	\end{subfigure}
	\hfill
    \begin{subfigure}[t]{0.49\textwidth}
        \centering
		\includegraphics[width=\textwidth]{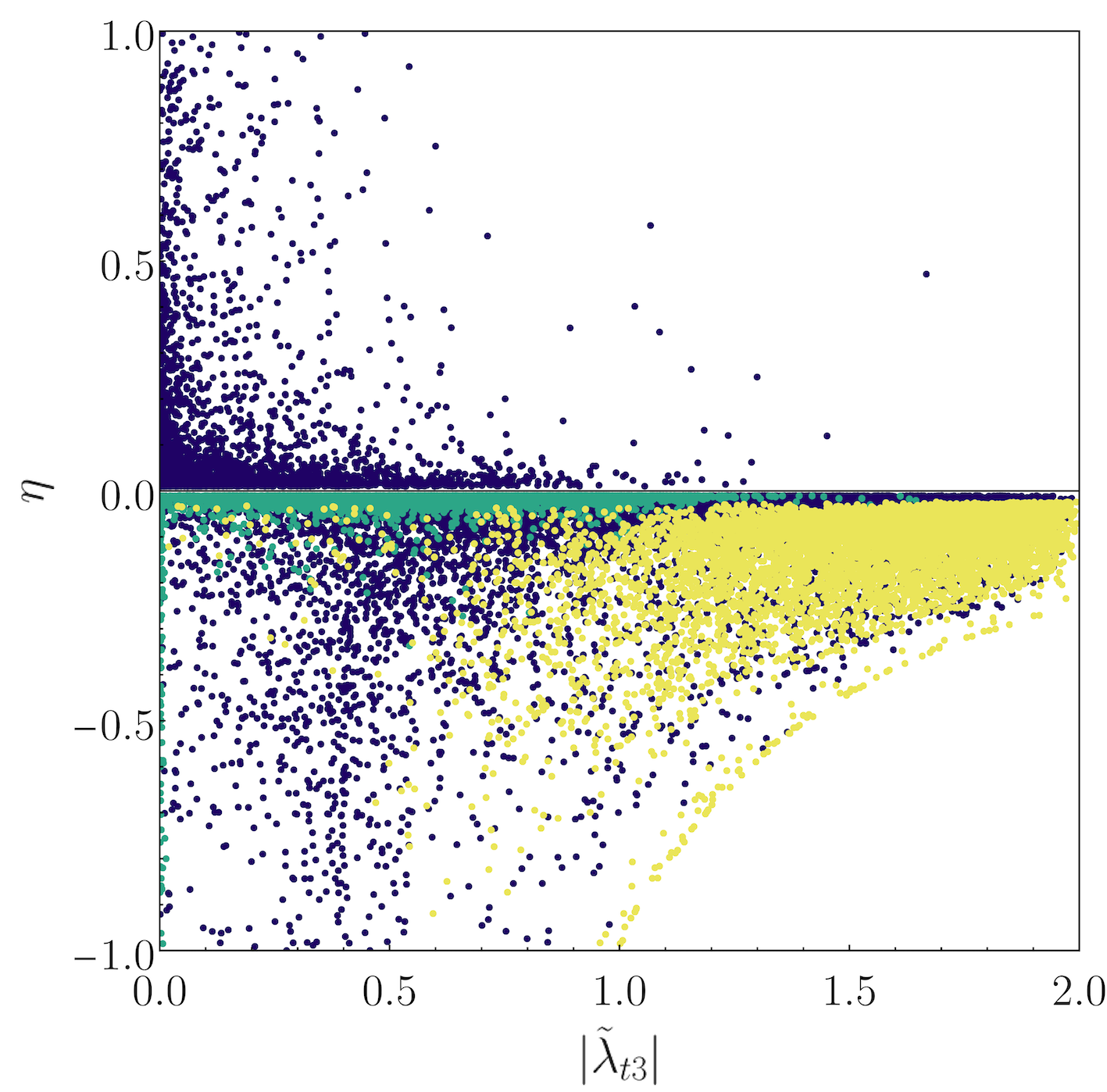}
        
        \label{fig:canonic:lt3eta}
	\end{subfigure}
	\caption{Viable parameter space for the canonical freeze-out scenarios. The yellow and green points correspond to the SFF and QDF scenarios respectively, while the blue points show the GCF scenario (excluding points belonging to either of the previous two).}
    \label{fig:canonic}
\end{figure}

The results of our scan for the canonical freeze-out scenarios are shown in Fig.~\ref{fig:canonic}. In yellow, green, and blue we display allowed points in the sub-scenarios SFF, QDF, and GCF, respectively. As anticipated, the GCF opens up allowed parameter space beyond the one found in Ref.~\cite{Acaroglu:2021qae}, in particular, for intermediate mass splittings $\Delta m_{13}$ (see upper left panel) and towards small $\Delta m_{\phi3}$ (see upper right panel). For these points, coannihilation effects with the heavier $\chi_i$ and $\phi$, respectively, become significant. Note however the existence of a completely unpopulated region for small $\Delta m_{\phi3}$ extending up to DM masses of about one TeV (lower left corner of the right panel). In this region, the effect of $\phi$ pair annihilation alone dilutes the DM number density below the measured value for efficient conversions, providing a solution within the conversion-driven freeze-out, see Sec.~\ref{sec:scan:conversion}. Note that the unpopulated regions in the upper right corners of the upper panels are simply a consequence of our chosen scan ranges. 

The lower panels in Fig.~\ref{fig:canonic} show the possible range of the DM mass splitting parameter $\eta$. While both the QDF and SFF scenario are by construction limited to $\eta <0$, the GCF scenario extends the allowed parameter space to arbitrary signs of $\eta$, \emph{i.e.}~within GCF it is possible that the lightest DM flavour $\chi_3$ has the weakest coupling to the SM\@. We also observe different allowed ranges for $\eta$ in the QDF vs.~SFF scenarios. In the QDF scenario, small values of $\eta$ are required to reproduce the quasi-degenerate DM spectrum, unless the DM coupling is very small. In the SFF scenario, on the other hand, larger values of $|\eta|$ and significant DM couplings are required to produce a significant splitting between the three DM generations. The latter is particularly true for the DM top coupling~$|\tilde\lambda_{t3}|$, \emph{i.e.}~we confirm the finding of Ref.~\cite{Acaroglu:2021qae} that DM in the SFF scenario is predominantly top-flavoured.

\subsection{Conversion-driven freeze-out scenario}
\label{sec:scan:conversion}

To generate points in the conversion-driven freeze-out scenario, we proceed in a way similar to the canonical freeze-out considered in Sec.~\ref{sec:scan:canonic}. For the sub-scenarios listed in Tab.~\ref{tab:nscheme}, we perform scans by randomly sampling the parameter space according to the scan ranges and priors listed in Tab.~\ref{tab:ranges:conversion}. However, since we use our own code to solve the coupled system of Boltzmann equations from Eqs.~\eqref{eq:BME_full_chi} and \eqref{eq:BME_full_phi}, the computation of the relic density is considerably more time-consuming than in the canonical scenario. At the same time, there is a complex interplay of input parameters that makes it non-trivial to satisfy the relic density constraint. As a result, in the case of the conversion-driven freeze-out scenario, rejecting all points that do not satisfy this condition appears to be very costly. Therefore, after obtaining a set of points that satisfy the flavour and (in)direct detection constraints, we proceed as follows.

\begin{table}[htb]
	\centering
    \caption[Scan ranges for the conversion driven freeze-out scenarios. The first column displays the parameter, the middle columns its scan range in the different scenarios and the right columns the respective prior. The upper part of the table shows the parameters contained in the coupling matrix $\tilde\lambda$, while the lower part shows the mass parameters.]{Scan ranges for the conversion driven freeze-out scenarios. The first column displays the parameter, the middle columns its scan range in the different scenarios and the right columns the respective prior. The upper part of the table shows the parameters contained in the coupling matrix $\tilde\lambda$, while the lower part shows the mass parameters.\footnotemark
    }
    \vspace{2mm}
	\begin{tabular}{ccccccc} 
    \toprule
    \multirow{2}{*}{\textbf{parameter}}& & \textbf{range} & & & \textbf{prior} & \\ 
    \cmidrule(lr){2-4}
    \cmidrule(lr){5-7}
     & C$_\chi 1_u$ & C$_\phi 1_u$ & C$_\phi 2_u$ & C$_\chi 1_u$ & C$_\phi 1_u$ & C$_\phi 2_u$ \\
    \bottomrule
    \toprule
	$D_1$ & $[10^{-7},10^{-5}]$ & $[10^{-7},10^{-5}]$ & $[10^{-7},10^{-5}]$ & log & log & log \\ 
    $D_2$ & $[0.1,2]$ & $[0.2,2]$ & $[10^{-7},10^{-5}]$ & lin & lin & log \\ 
    $D_3$ & $[0.1,2]$ & $[0.2,2]$ & $[0.2,2]$ & lin & lin & log \\ 
    $\theta_{12}$& $[10^{-8},10^{-5}]$ & $[10^{-9},10^{-7}]$ & $[0,\pi/4]$ & log & log & lin \\
    $\theta_{13}$& $[10^{-8},10^{-5}]$ & $[10^{-9},10^{-7}]$ & $[10^{-10},10^{-8}]$ & log & log & log \\
    $\theta_{23}$ & $[0,\pi/4]$ & $[0,\pi/4]$ & $[10^{-10},10^{-8}]$ & lin & lin & log\\
    $\phi_{ij}$& $0$ & $0$ & $0$ & -- & -- & -- \\
    $\delta_{ij}$& $[0,2\pi)$ & $[0,2\pi)$ & $[0,2\pi)$ & lin & lin & lin \\
    $\gamma_i$& $[0,2\pi)$ & $[0,2\pi)$ & $[0,2\pi)$ & lin & lin & lin \\ \midrule
    $\eta$& $[0.005,0.1]$ & $[0.5,1]$ & $[0.5,1]$ & lin & lin & lin \\ 
    $m_\chi\,[\mathrm{GeV}]$ & $[100,2000]$ & $[100,2000]$ & $[100,2000]$ & lin & lin & lin \\
    $m_\phi\,[\mathrm{GeV}]$ & $[1.2 \, m_{\chi_1},3 \, m_{\chi_1}]$ & $[m_{\chi_3},1.1 \, m_{\chi_3}]$ & $[m_{\chi_2},1.1 \, m_{\chi_2}]$ & lin & lin & lin \\
    \bottomrule
	\end{tabular}
	\label{tab:ranges:conversion}
\end{table}

As a first step, we compute the relic density using \textsc{micrOMEGAs}, (falsely) assuming chemical equilibrium among all $\mathbbm{Z}_2$-odd particles. Since inefficient conversion rates will only increase the DM density in the targeted region of parameter space, we select all points that appear underabundant in the above approximation, as these points provide potential solutions via conversion-driven freeze-out.  Specifically, we require that the computed $\Omega h^2$ is at least 30\% below the measured value. In a second step, for each of the selected points, we vary the respective very weak coupling(s) ($D_1$ in the scenarios C$_\chi1_u$ and C$_\phi1_u$, and $D_1, D_2$ in the scenario C$_\phi2_u$) until the relic density is matched. To do this, we use a minimisation algorithm. In this procedure, all other parameters of a given point are left untouched. Furthermore, in the case of C$_\phi2_u$, the ratio $D_1/D_2$ obtained in the initial sampling is maintained so that both couplings are rescaled by the same factor. Note that the scan ranges defined in Tab.~\ref{tab:ranges:conversion} do not apply during this rescaling. 

\begin{figure}
    \centering    \includegraphics[width=0.48\textwidth]{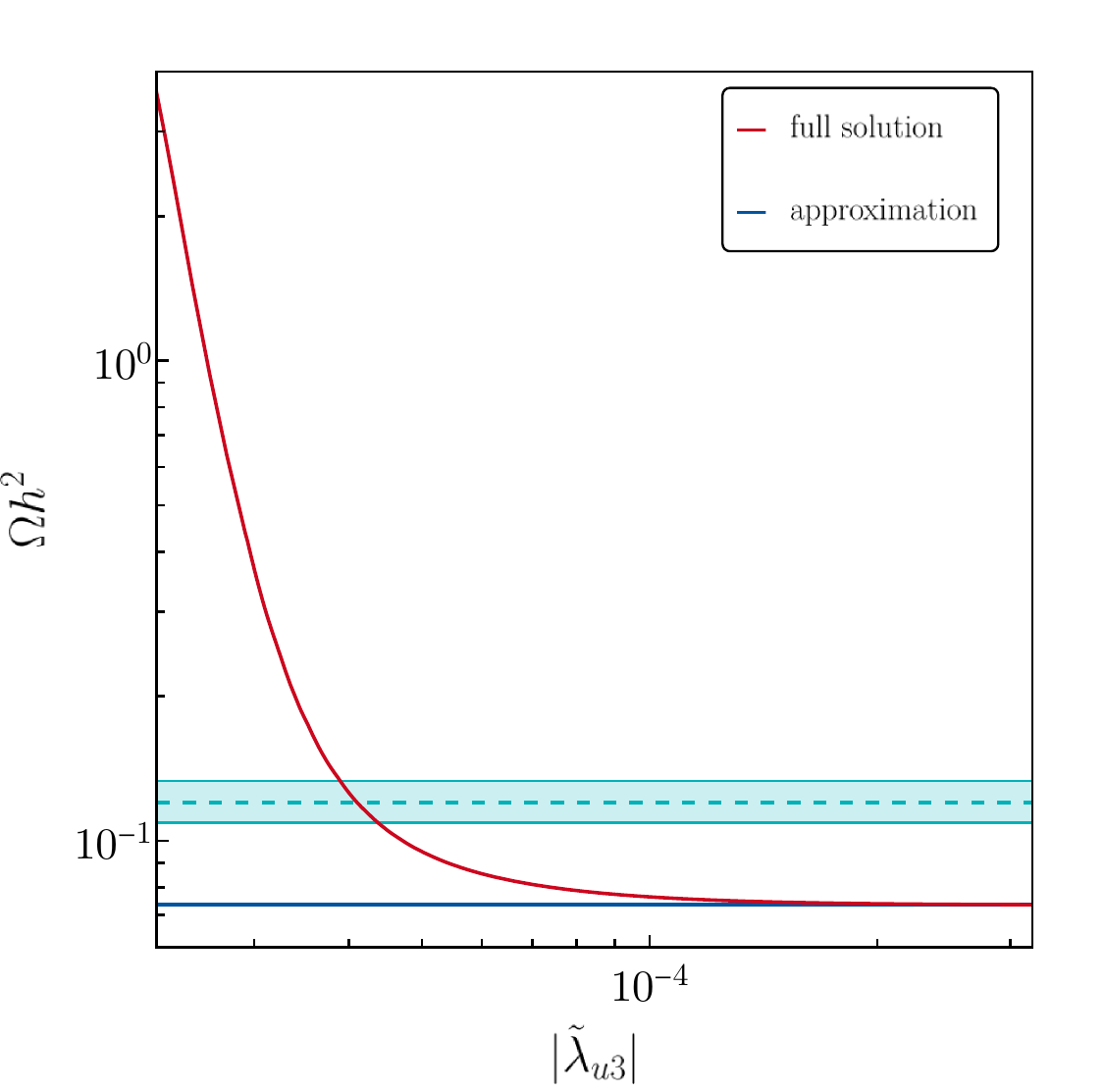}
    \includegraphics[width=0.48\textwidth]{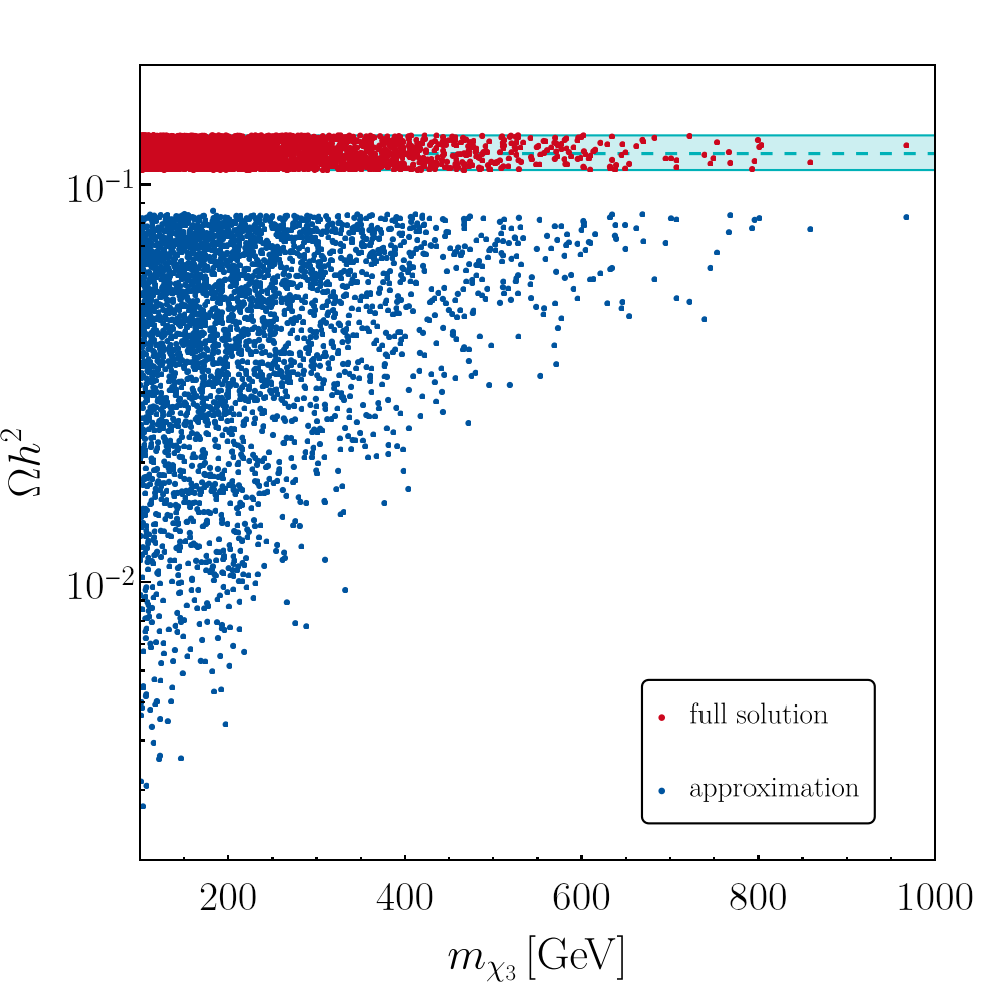}

    \caption{Determination of the DM coupling by varying the coupling until the correct relic abundance is obtained within a 10\% uncertainty range. Left panel: Relic abundance computed with the approximation and full Boltzmann equations when the coupling of $\chi_3$ to up-quarks is varied. Right panel: Rescaled scan for C$_\chi 1_u$.}
    \label{fig:rescale_chi_u1}
\end{figure}

Figure~\ref{fig:rescale_chi_u1} illustrates the procedure using the C$_\chi1_u$ scenario as an example. In the left panel we show the relic density as a function of the physical coupling $|\tilde \lambda_{u3}|$ for an exemplary point selected in our sampling within the scenario. The blue line shows the value obtained by assuming chemical equilibrium, which is independent of the coupling in the range shown. According to our selection criterion, the DM relic density is below the measured value indicated by the cyan dashed line. The red curve shows the full solution, which rises steeply towards lower coupling due to lower conversion rates. Our minimisation algorithm finds the coupling that gives a value of $\Omega h^2$ within the cyan shaded band, \emph{i.e.}~within the assumed theoretical uncertainty around the measured value. In the case shown, it finds $|\tilde \lambda_{u3}| \simeq 4 \times 10^{-5}$.
The right panel shows the full set of points in the C$_\chi1_u$ scenario in the chemical equilibrium approximation (blue points) and after the rescaling algorithm using the solution of the coupled set of Boltzmann equations (red points). The couplings found are in a relative wide range, between $10^{-7}$ and several times $10^{-4}$, in this scenario. For the scenarios C$_\phi 1_u$ and C$_\phi 2_u$ the very weak couplings are typically around a few times $10^{-7}$. Solutions of the Boltzmann equations as a function of $x$ are shown in App.~\ref{qpp:abund} for two exemplary points.

\begin{figure}
	\centering
	\begin{subfigure}[t]{0.48\textwidth}
        \centering
		\includegraphics[width=\textwidth]{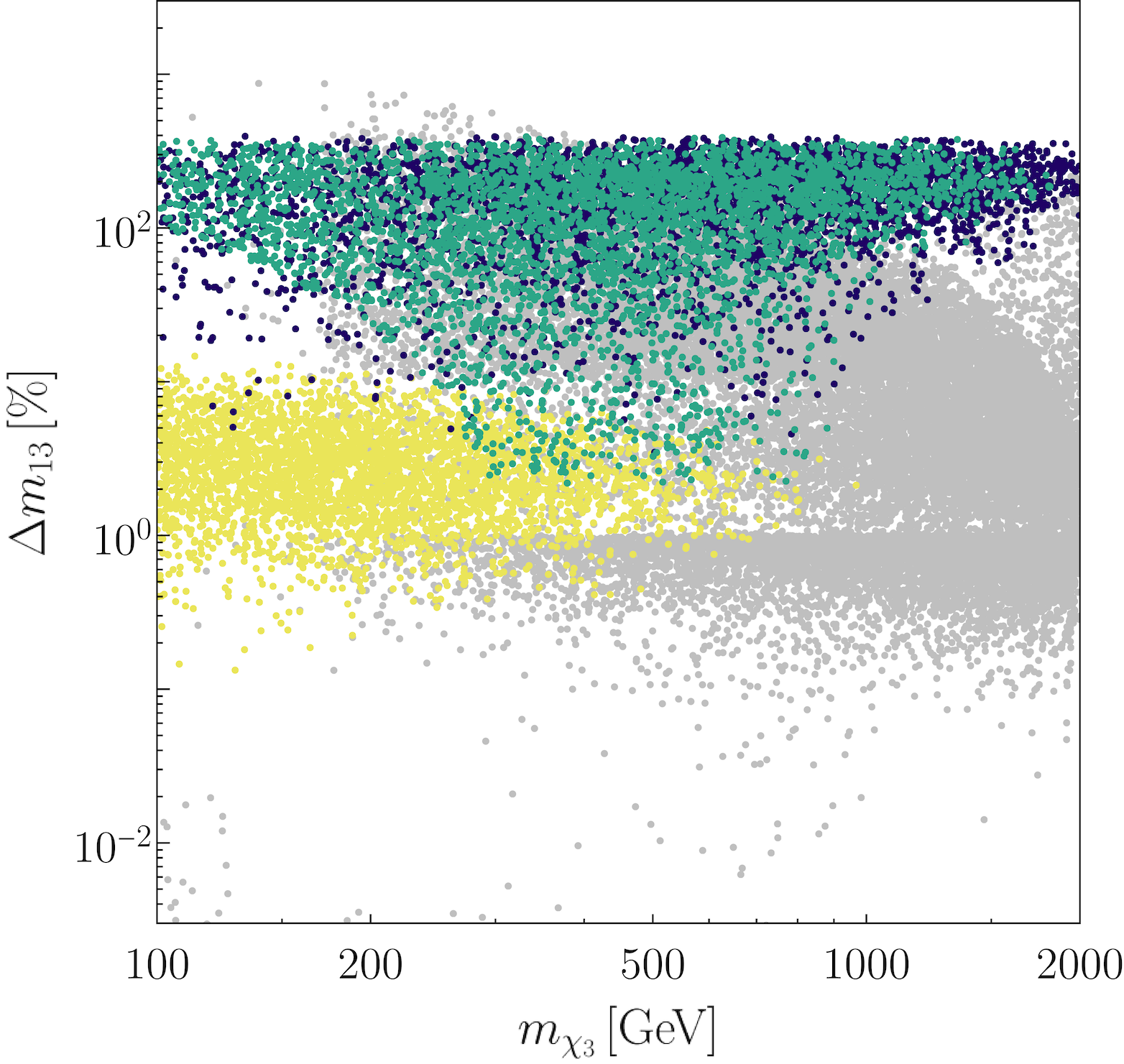}
        \label{fig:cdfo:mphimchi3}
	\end{subfigure}
	\hfill
        \begin{subfigure}[t]{0.48\textwidth}
        \centering
		\includegraphics[width=\textwidth]{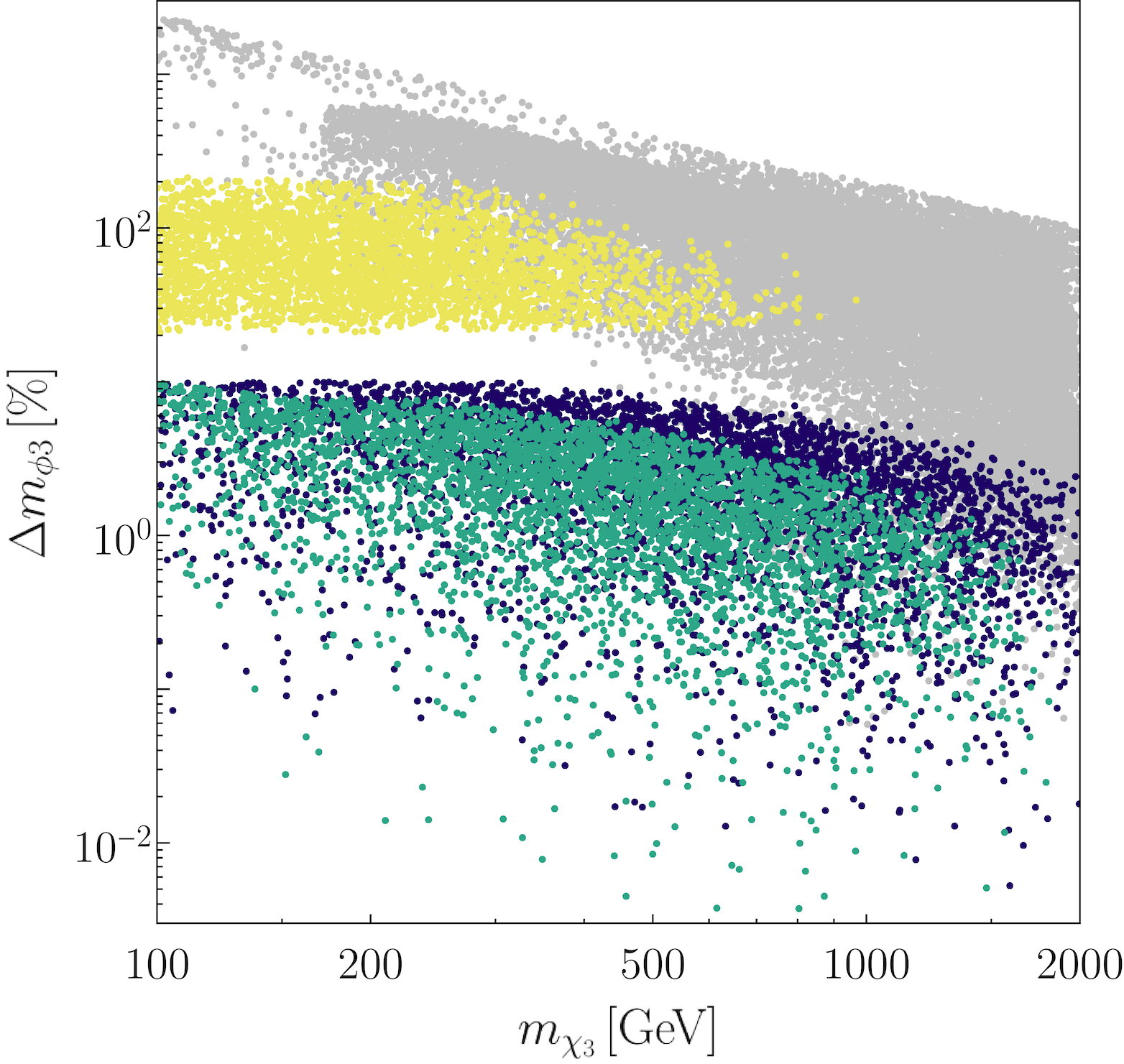}
        \label{fig:cdfo:mchi3Dm13}
	\end{subfigure}
	\caption{Same as Fig.~\ref{fig:canonic} but for the conversion-driven freeze-out scenarios. Specifically, the yellow, green and blue points correspond to the allowed points in the C$_\chi1_u$, C$_\phi2_u$ and C$_\phi1_u$ scenarios, respectively. For comparison, the light grey points show the full set of allowed points in the canonical scenario shown in Fig.~\ref{fig:canonic}.}
    \label{fig:cdfo}
\end{figure}

\footnotetext{Note that we set the mixing angles $\phi_{ij}$ of the orthogonal matrix $O$ to zero for simplicity. Non-zero values would only dilute the splitting between different couplings and thus not induce any qualitatively new effect.}

In Fig.~\ref{fig:cdfo}, we present the results for the conversion-driven freeze-out scenarios. In yellow, blue and green we denote the parameter points in the sub-scenarios C$_\chi 1_u$, C$_\phi 1_u$ and C$_\phi 2_u$, respectively, which are allowed by all constraints considered so far, \emph{i.e.}~flavour, (in)direct detection and relic density constraints. For comparison, the allowed parameter points of the canonical freeze-out scenario (including the sub-scenarios SFF and QDF) are shown in grey. 
The left and right panels of Fig.~\ref{fig:cdfo} show the DM mass versus the relative mass splittings $\Delta m_{13}=(m_{\chi_1}-m_{\chi_3})/m_{\chi_3}$ and $\Delta m_{\phi3}=(m_\phi-m_{\chi_3})/m_{\chi_3}$, respectively. In the right panel, the sub-scenarios C$_\phi 1_u$ and C$_\phi 2_u$ populate the lower left region, where there are no points in the canonical scenario, because DM is underabundant for efficient conversions due to the large QCD contribution to the $\phi$ pair annihilation cross section (which depends only on $m_\phi$). In this respect, our model shares the same phenomenology as the simplified model considered in Ref.~\cite{Garny:2017rxs} with only one Majorana fermion $\chi$ (analogous results have also been obtained for other spin assignments with minimal particle content~\cite{Arina:2023msd}). However, unlike the minimal model of Ref.~\cite{Garny:2017rxs}, which has a sharp boundary separating the canonical and conversion-driven freeze-out regions, we observe an overlap of these regions, especially towards large $m_{\chi_3}$ and intermediate mass splittings. This is due to contributions to $\phi$-pair annihilation from $t$-channel diagrams involving $\chi_1$ and/or $\chi_2$, which can significantly increase the cross section if the respective couplings are large. A larger $\phi$-pair annihilation cross section pushes the conversion-driven freeze-out region towards larger masses (and larger mass splittings).
Comparing C$_\phi 1_u$ and C$_\phi 2_u$, we find that the former scenario allows for somewhat larger mass splittings.
This can be explained as follows. In C$_\phi 2_u$, the number of degrees of freedom lighter than $\phi$ is doubled. These states can only deplete through conversion processes into $\phi$. At the same time, the total rate of dark sector depletion is limited to that of $\phi$-pair annihilation. Hence, for a parameter point in C$_\phi 2_u$, the resulting number density in the very weakly interacting sector is larger than for a similar point but with $\chi_2$ considerably heavier than $\phi$ (which would belong to C$_\phi 1_u$). Therefore, for a given DM mass, the largest mass splittings that still provide viable solutions in C$_\phi 1_u$ would overshoot the relic density (even for efficient conversions) in C$_\phi 2_u$ requiring a smaller mass splitting effectively enhancing the dilution rate in the dark sector.  

In the sub-scenario C$_\chi 1_u$, we require $m_\phi$ to be at least 20\% larger than $m_{\chi_1}$ such that $\chi_1,\chi_2$ provide the dominant contribution to the overall annihilation rate (given their sizeable couplings). Hence, the sampled points of the C$_\chi 1_u$ scenario populate the upper region of the right panel of Fig.~\ref{fig:cdfo}, while being characterised by small mass splittings among the $\chi_2$, see left panel. Note that the region of small $\Delta m_{13}$ is not forbidden in the canonical freeze-out scenario (unlike the region of small $\Delta m_{\phi3}$), since the $\chi_i$ pair annihilation cross section can be arbitrarily small for arbitrarily small couplings. For this reason, the canonical freeze-out scenario and C$_\chi 1_u$ overlap in a large part of the parameter space. Note that solutions in the C$_\chi 1_u$ sub-scenario favour relatively small (positive) values of $\eta$ for the following reason. First, the mass splitting $\Delta m_{13}$ must be small to maximise coannihilation effects. At the same time, the annihilation channel of the heavier states $\chi_i$ must be large, requiring large couplings of $\chi_1$ and/or $\chi_2$. According to Eq.~\eqref{eq:masscorr}, these two requirements can only be fulfilled for small $\eta$. 

Let us briefly comment on our decision to consider scenarios with a dominant DM coupling to the $u$ quark by briefly discussing the case of a dominant coupling to the top quark listed in Tab.~\ref{tab:nscheme}. It is well known that $\chi_i\chi_i$ annihilation into a fermion pair is helicity suppressed. Accordingly, annihilation into top pairs provides by far the largest cross section. Therefore, in order to achieve a sufficiently large annihilation cross section of the heavier states $\chi_1$ and/or $\chi_2$ needed for a successful conversion-driven freeze-out in the mass range of interest, these states must couple dominantly to the top. By definition, this is not the case in C$_\chi 1_t$. Therefore, its cosmologically allowed parameter space is significantly reduced. This is also true for C$_\phi 1_t$, but for different reasons. According to Eq.~\eqref{eq:masscorr}, the sub-scenario C$_\phi 1_t$ requires large couplings for $\chi_1$ and $\chi_2$ to achieve a large mass splitting between these states and $\chi_3$. However, large couplings are disfavoured by flavour constraints if $\chi_1$ and $\chi_2$ couple only to light quarks, leading to very few allowed points in the parameter space. Note that in both scenarios, C$_\chi 1_t$ and C$_\phi 1_t$, the appearance of the massive top in the conversion processes leads to a further suppression of the respective rates, allowing slightly larger DM couplings, see \emph{e.g.}~Ref.~\cite{Garny:2018icg}.
Finally, in analogy to the case of C$_\phi 2_u$, in the case of C$_\chi 2_u$ the allowed conversion-driven freeze-out region is reduced, reaching down to smaller values of $m_{\chi_3}$ compared to C$_\chi 1_u$ and is therefore not presented in detail here. 

Finally, we comment on the effect of two simplifications made in our calculation of the relic density. First, the coupled set of Boltzmann equations~\eqref{eq:BME_full_chi} and \eqref{eq:BME_full_phi}, implicitly assume the maintenance of kinetic equilibrium throughout the freeze-out process. In the preferred coupling range of conversion-driven freeze-out, this assumption usually cannot be justified by efficient elastic scattering of DM with the thermal bath. However, it has been shown that due to the small relative mass splitting, DM inherits a (nearly) thermal distribution of the annihilating particle, justifying the use of the integrated Boltzmann equation with a correction of the order of $\lesssim 10\%$~\cite{Garny:2017rxs}.
Another relevant correction comes from non-relativistic effects of the coloured mediator $\phi$, in particular Sommerfeld enhancement and bound state formation effects. As shown in Ref.~\cite{Garny:2021qsr}, the latter in particular can be large and significantly increase the viable parameter space within the conversion-driven freeze-out (see also Ref.~\cite{Binder:2023ckj} for a refined description of excited  bound states, which are important in this scenario). However, the inclusion of these effects in our model is beyond the scope of this work and is left for future investigation. It is expected to extend the conversion-driven freeze-out region in the sub-scenarios C$_\phi 1_u$ and C$_\phi 2_u$ towards larger masses, while leaving the qualitative results unchanged.

\section{LHC signatures}
\label{sec:LHC}

Our model of flavoured Majorana DM, like all models with coloured $t$-channel mediators~\cite{Papucci:2014iwa}, is severely constrained by LHC searches, in particular by the production and decay of the mediator $\phi$. Therefore, in Ref.~\cite{Acaroglu:2021qae} a simple recast of the most constraining LHC search for supersymmetric squarks~\cite{CMS:2019zmd} was performed. It was found that the $t$-channel exchange of the Majorana DM provides a significant contribution to the production of mediator pairs, which is strongly enhanced by the up-quark parton distribution function for the $uu\to\phi\phi$ process. It was pointed out that the same-sign process $pp\to tt+\slashed{E}_T$ is predicted at significant rates, a signature for which no dedicated searches exist.

In this section we revisit the LHC signatures and the resulting available constraints on the parameter space of our model. First, we check the compatibility of the viable parameter points found in the previous section with current LHC data by comparing them with the exclusion limits from a large number of searches using the \textsc{SModelS}~\cite{Alguero:2021dig,MahdiAltakach:2023bdn} reinterpretation tool. This exercise also allows us to identify promising future directions for discovering the new particles predicted by our model at the LHC. We then turn our attention to the model-specific signatures of flavoured Majorana DM implied by the aforementioned Majorana-specific $t$-channel same-sign contribution.

\subsection{LHC simplified  model constraints} \label{sec:SMSconst}

As a first check on the compatibility of the viable parameter points with current LHC data, we consider exclusion limits from a large number of searches using the \textsc{SModelS} package~\cite{Alguero:2021dig,MahdiAltakach:2023bdn}. This reinterpretation tool contains a large number of LHC results interpreted for simplified models of NP\@. We run \textsc{SModelS} using the corresponding interface of \textsc{micrOMEGAs}~\cite{Barducci:2016pcb} to calculate the LHC production cross sections and decay tables. With this tool chain we derive the constraints from LHC searches on our model within the different scenarios considered in Sec.~\ref{sec:scans}. A global overview of all relevant searches excluding points from any of these scenarios is given in Tab.~\ref{tab:lhcsearches}.

\begin{table}
	\centering
    \caption{Relevant LHC searches for each scenario identified in our \textsc{SmodelS}~\cite{Alguero:2021dig,MahdiAltakach:2023bdn} analysis. The first column contains the scenario, the second the search report number, the third the center-of-mass energy of the LHC run the respective data set is based on, and the fourth column contains the relevant signature. }
    \vspace{2mm}
        \setlength{\tabcolsep}{8pt}
	\begin{tabular}{cccc} 
    \toprule
    \textbf{scenario}& \textbf{search}& \boldmath$\sqrt{s}$ & \textbf{signatures} \\ 
    \bottomrule 
    \toprule
\multirow{11}{*}{{canonic}}& \texttt{ATLAS-SUSY-2013-02}~\cite{ATLAS:2014jxt} & $8\,\mathrm{TeV}$ & jets+$\slashed{E}_T$ \\ 
& \texttt{ATLAS-SUSY-2016-07}~\cite{ATLAS:2017mjy} & $13\,\mathrm{TeV}$ & jets+$\slashed{E}_T$ \\ 
& \texttt{ATLAS-SUSY-2016-15}~\cite{ATLAS:2017drc} & $13\,\mathrm{TeV}$ & tops+$\slashed{E}_T$ \\ 
& \texttt{ATLAS-SUSY-2018-12}~\cite{ATLAS:2020dsf} & $13\,\mathrm{TeV}$ & tops+$\slashed{E}_T$ \\ 
& \texttt{ATLAS-SUSY-2018-22}~\cite{ATLAS:2020syg} & $13\,\mathrm{TeV}$ & jets+$\slashed{E}_T$ \\ 
& \texttt{CMS-SUS-16-033}~\cite{CMS:2017abv} & $13\,\mathrm{TeV}$ & jets+$\slashed{E}_T$ \\ 
& \texttt{CMS-SUS-16-036}~\cite{CMS:2017okm} & $13\,\mathrm{TeV}$ & jets+$\slashed{E}_T$ \\ 
& \texttt{CMS-SUS-19-006}~\cite{CMS:2019zmd} & $13\,\mathrm{TeV}$ & jets+$\slashed{E}_T$ \\ 
& \texttt{CMS-SUS-19-009}~\cite{CMS:2019ysk} & $13\,\mathrm{TeV}$ & tops+$\slashed{E}_T$ \\ 
& \texttt{CMS-SUS-20-002}~\cite{CMS:2021eha} & $13\,\mathrm{TeV}$ & tops+$\slashed{E}_T$ \\ 
\midrule
\multirow{11}{*}{{C$_\chi 1_u$}} &\texttt{ATLAS-SUSY-2013-02}~\cite{ATLAS:2014jxt} & $8\,\mathrm{TeV}$ & jets+$\slashed{E}_T$ \\ 
& \texttt{ATLAS-SUSY-2013-21}~\cite{ATLAS:2014hqe} & $8\,\mathrm{TeV}$ & $cc+\slashed{E}_T$ \\ 
& \texttt{ATLAS-SUSY-2014-03}~\cite{ATLAS:2015jct} & $8\,\mathrm{TeV}$ & $cc+\slashed{E}_T$ \\ 
& \texttt{ATLAS-SUSY-2016-07}~\cite{ATLAS:2017mjy} & $13\,\mathrm{TeV}$ & jets+$\slashed{E}_T$ \\ 
& \texttt{ATLAS-SUSY-2016-15}~\cite{ATLAS:2017drc} & $13\,\mathrm{TeV}$ & tops+$\slashed{E}_T$ \\ 
& \texttt{ATLAS-SUSY-2016-26}~\cite{ATLAS:2018qzb} & $13\,\mathrm{TeV}$ & $cc+\slashed{E}_T$ \\ 
& \texttt{CMS-SUS-16-033}~\cite{CMS:2017abv} & $13\,\mathrm{TeV}$ & jets+$\slashed{E}_T$ \\ 
& \texttt{CMS-SUS-16-036}~\cite{CMS:2017okm} & $13\,\mathrm{TeV}$ & jets+$\slashed{E}_T$ \\ 
& \texttt{CMS-SUS-16-049}~\cite{CMS:2017mbm} & $13\,\mathrm{TeV}$ & tops+$\slashed{E}_T$ \\ 
& \texttt{CMS-SUS-19-006}~\cite{CMS:2019zmd} & $13\,\mathrm{TeV}$ & jets+$\slashed{E}_T$ \\ 
& \texttt{CMS-SUS-20-002}~\cite{CMS:2021eha} & $13\,\mathrm{TeV}$ & tops+$\slashed{E}_T$ \\ 
\midrule
\multirow{2}{*}{{C$_\phi 1_u$}}  &\texttt{ATLAS-SUSY-2016-32}~\cite{ATLAS:2019gqq} & $13\,\mathrm{TeV}$ & stable R-hadron \\ 
& \texttt{CMS-PAS-EXO-16-036}~\cite{CMS-PAS-EXO-16-036} & $13\,\mathrm{TeV}$ & stable R-hadron \\
\midrule 
\multirow{6}{*}{{C$_\phi 2_u$}} &\texttt{ATLAS-SUSY-2013-02}~\cite{ATLAS:2014jxt} & $8\,\mathrm{TeV}$ & jets+$\slashed{E}_T$ \\ 
& \texttt{ATLAS-SUSY-2016-32}~\cite{ATLAS:2019gqq} & $13\,\mathrm{TeV}$ & stable R-hadron \\
& \texttt{CMS-PAS-EXO-16-036}~\cite{CMS-PAS-EXO-16-036} & $13\,\mathrm{TeV}$ & stable R-hadron \\
& \texttt{CMS-SUS-16-032}~\cite{CMS:2017kil} & $13\,\mathrm{TeV}$ & $cc+\slashed{E}_T$ \\ 
& \texttt{CMS-SUS-16-036}~\cite{CMS:2017okm} & $13\,\mathrm{TeV}$ & jets+$\slashed{E}_T$ \\ 
& \texttt{CMS-SUS-16-049}~\cite{CMS:2017mbm} & $13\,\mathrm{TeV}$ & tops+$\slashed{E}_T$ \\ 
    \bottomrule
	\end{tabular}
	\label{tab:lhcsearches}
\end{table}

\begin{figure}
	\centering
		\begin{subfigure}[t]{0.49\textwidth}
		\includegraphics[width=\textwidth]{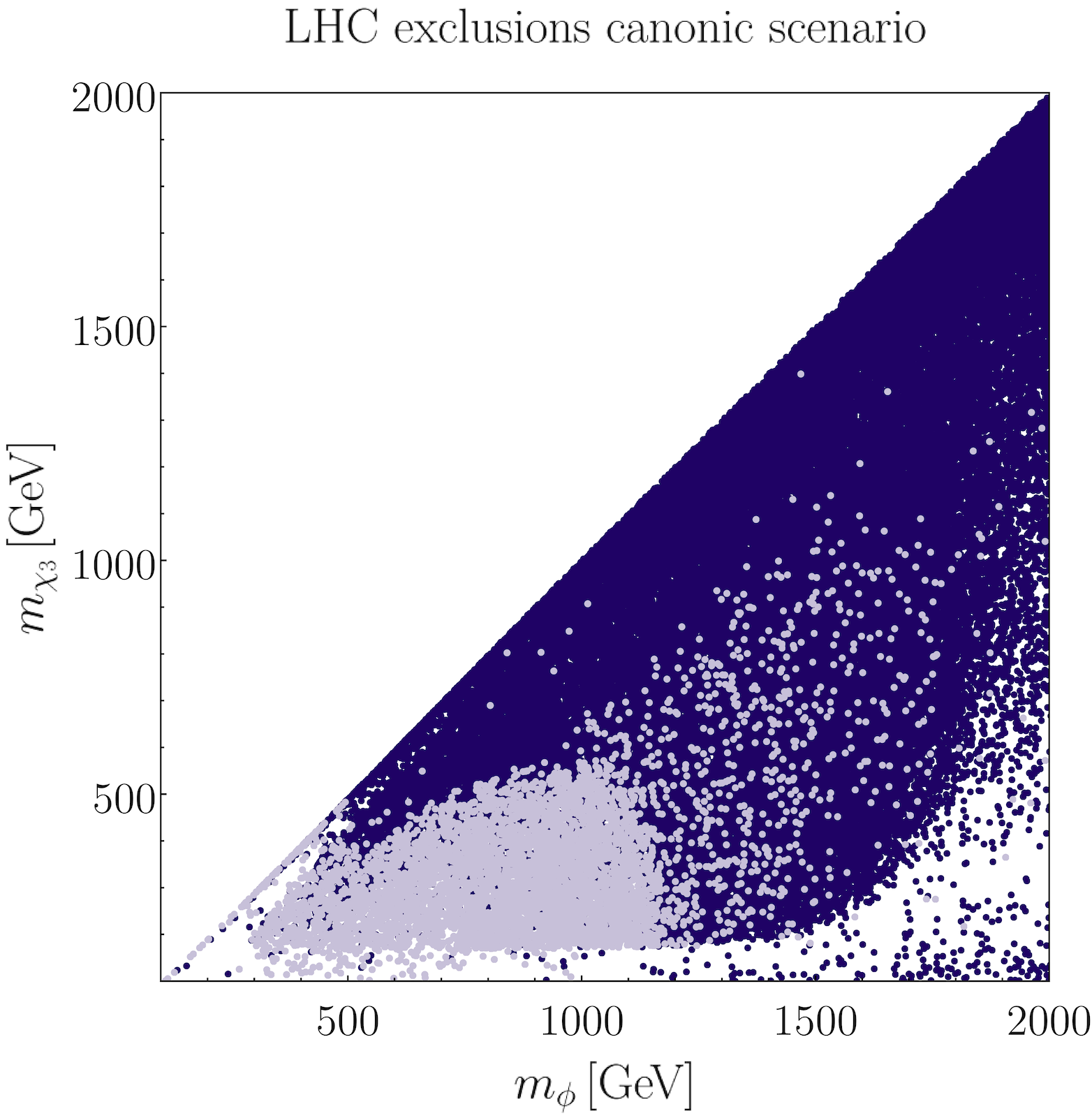}
		\end{subfigure}
  	\begin{subfigure}[t]{0.49\textwidth}
		\includegraphics[width=\textwidth]{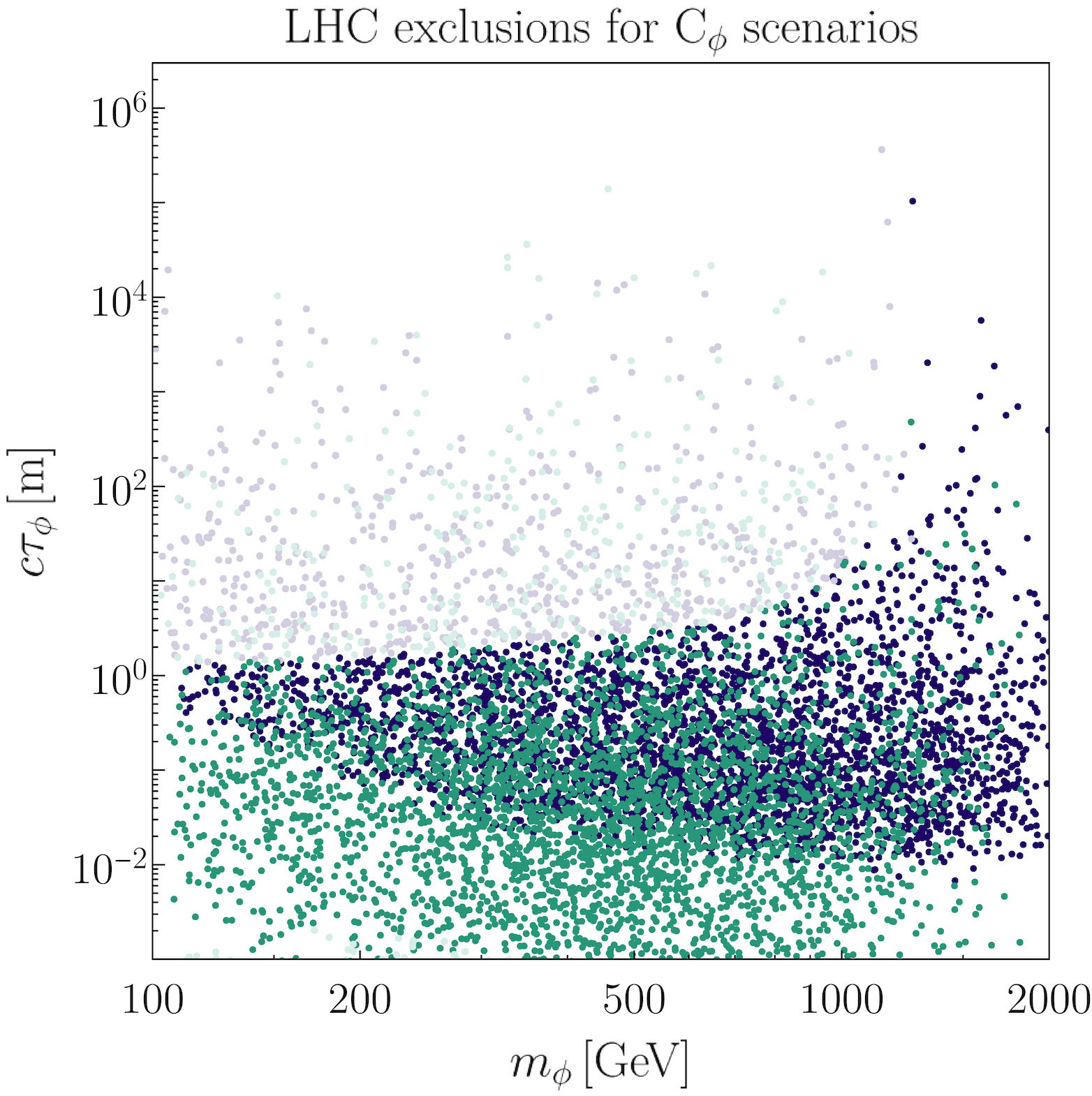}
		\end{subfigure}
		\begin{subfigure}[t]{0.49\textwidth}
		\includegraphics[width=\textwidth]{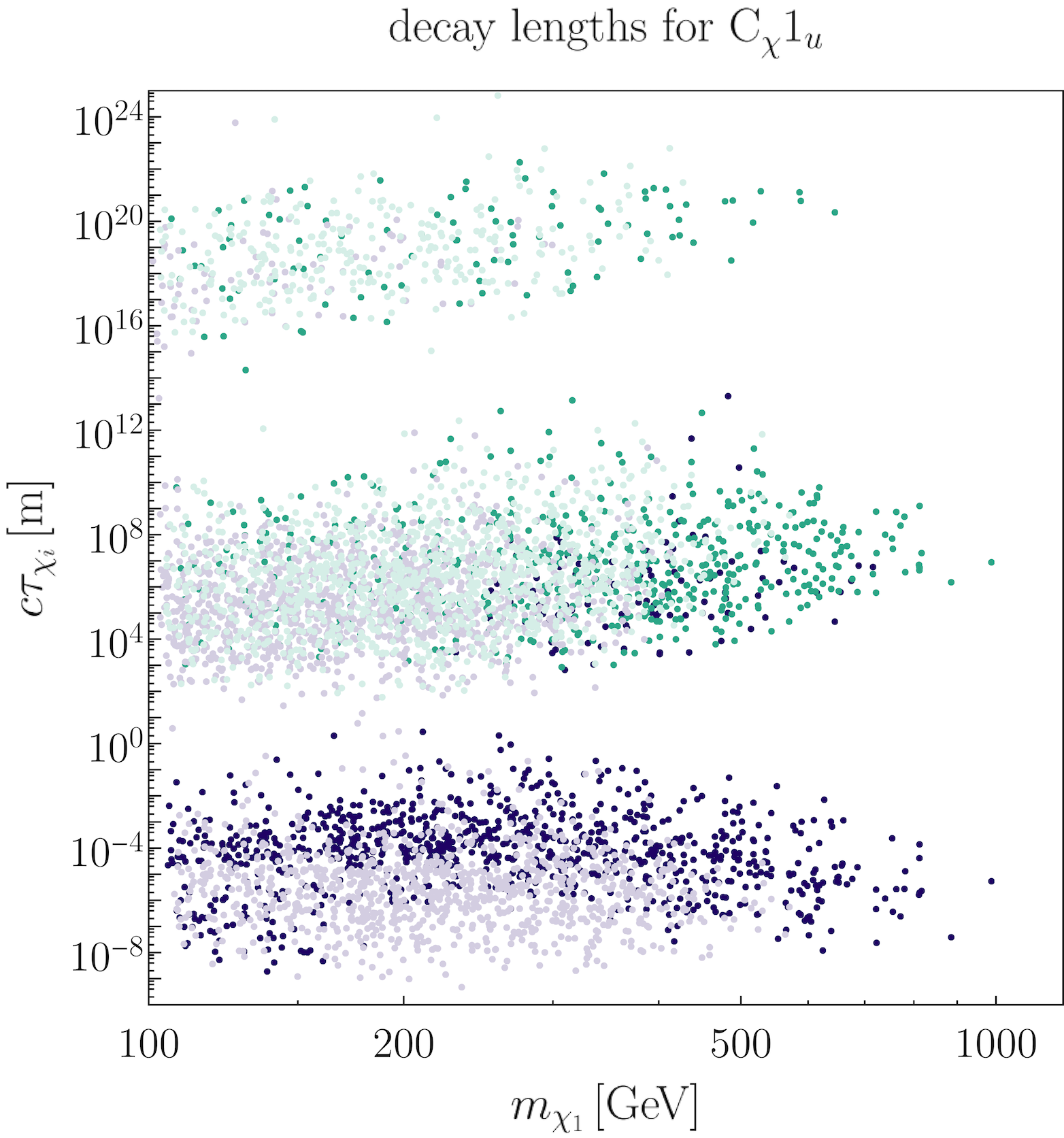}
		\end{subfigure}
        \caption{Upper left panel: LHC constraints from $\slashed E_T$ searches within the canonical freeze-out scenario (left) and the conversion-driven freeze-out scenario C$_\chi1_u$ in the $m_\phi$-$m_{\chi_3}$ mass plane. The dark blue (light shaded) points are allowed (excluded at 95\% CL). Upper right panel: LHC constraints from long-lived particle searches in the conversion-driven freeze-out scenarios C$_\phi1_u$ (blue) and C$_\phi2_u$ (green) in the plane spanned by the mass and decay length of $\phi$. The light shaded areas are excluded at 95\% CL\@. Lower panel: Decay lengths of $\chi_1$ (blue) and $\chi_2$ (green) versus $m_{\chi_1}$ in the C$_\chi1_u$ scenario.}
        \label{fig:collider:scan}
\end{figure}

The results of this analysis are shown in Fig.~\ref{fig:collider:scan}. In the upper left panel we show the effect of the LHC constraints in the $m_\phi$--$m_{\chi_3}$ plane for the canonical freeze-out scenario. Note that here we do not distinguish between the QDF, SFF and GCF scenarios defined in Sec.~\ref{sec:scan:canonic}, considering the union of the three sets of points (passing all non-LHC constraints).  The points excluded by LHC searches at 95$\%$ CL are shown in light shading, while the allowed points are shown in saturated colour (here dark blue). For a large fraction of the points, the exclusion contour is very similar to that obtained for the simplified models used in the original interpretation of the $\slashed E_T$ searches by the LHC collaborations, see references in Tab.~\ref{tab:lhcsearches}. This exclusion extends to mediator and DM masses of $m_\phi \simeq 1200\,\mathrm{GeV}$ and $m_{\chi_3}\simeq 600\,\mathrm{GeV}$, respectively. Similar results have been found for this model in~\cite{Acaroglu:2021qae}. However, in our scan there are several allowed points inside the exclusion contour described above, as well as a number of excluded points outside the exclusion contour. 

The allowed points with small DM and mediator masses inside the usual exclusion contour are characterised by very small mass splittings within the DM multiplet, $\chi_i$, such that (at least) one of the heavier states becomes metastable. Since the corresponding lifetime falls between the prompt and detector stable regime, it is \emph{a priori} not clear whether the respective $\slashed E_T$ searches apply here, and thus \textsc{SModelS} conservatively discards the signature. However, if the acceptance for non-prompt decays is known, a dedicated reinterpretation of the considered $\slashed E_T$ searches may exclude (some of) these points.

On the other hand, the excluded points at large masses, which lie outside the usual exclusion contour, all show an enhanced production cross section of $\phi$ pairs of the same sign via $\chi_i$ in the $t$-channel. This process is enhanced by the large couplings of the $\chi_i$ involved as well as the large up-quark parton densities in the relevant momentum range, see \emph{e.g.}~\cite{Garny:2013ama}. Here, $\phi$ subsequently decays into a pair of top quarks or jets. In both cases the total production cross section is strongly enhanced, as already found in Ref.~\cite{Acaroglu:2021qae}. Since the same-sign pair-production of $\phi$ is proportional to the Majorana mass $m_{\chi_i}$, it becomes increasingly relevant for large DM masses, in contrast to the production of opposite-sign final states. 

In summary, our analysis reveals a large number of model points between the electroweak and the TeV scale that are still allowed despite the large number of LHC searches. At the same time, the large associated production cross sections should provide an opportunity to probe these models with upcoming LHC searches.

We now discuss the LHC constraints for the conversion-driven freeze-out scenario.  In the sub-scenarios C$_\phi 1_u$ and C$_\phi 2_u$, $\phi$ can only decay via the very weak interaction(s) and is therefore rendered metastable, making the search for long-lived particles relevant. 
Therefore, in the upper right panel of Fig.~\ref{fig:collider:scan} we show our results in the plane spanned by the mediator mass and its decay length. The sub-scenarios C$_\phi 1_u$ and C$_\phi 2_u$ are marked in blue and green, respectively. Similar to what was found for the minimal model considered in Ref.~\cite{Garny:2017rxs}, the typical decay lengths are in the range of millimetres to several metres, while scattered points with significantly larger decay lengths exist as well. These are characterised by an extremely small mass splitting between DM and $\phi$. In our scan, the minimal decay length in C$_\phi 1_u$ is somewhat larger than in C$_\phi 2_u$. 
The light shading denotes the 95\% CL exclusion limit from the current long-lived particle searches at the LHC using \textsc{SModelS}. For C$_\phi 1_u$, the exclusion comes entirely from searches for detector-stable R-hadrons, which yield highly ionised tracks and anomalous time-of-flight. These searches exclude masses up to about 1.3\,TeV for the largest lifetimes, but become less sensitive to small decay lengths due to the suppressed number of mediators traversing the necessary parts of the detector. Finally, below about one metre, their constraining power is gone. Note that the searches for displaced jets and disappearing tracks make use of the decay within the detector and gain sensitivity for decay lengths (well) below a metre. However, their applicability to the model under consideration is questionable (see discussions in Ref.~\cite{Brooijmans:2020yij} on this topic) and therefore not considered by \textsc{SModelS}. In addition to R-hadron searches, C$_\phi 2_u$ is constrained by searches for $\slashed E_T$, see Tab.~\ref{tab:lhcsearches}.

Finally, in the lower panel of Fig.~\ref{fig:collider:scan} we show the results for the sub-scenario C$_\chi 1_u$ in the plane spanned by $m_{\chi_1}$ and the decay length of $\chi_1$ (blue) and $\chi_2$ (green). The $\chi_2$ state is always long-lived with decay lengths that typically exceed the length scales of the detector considerably, while $\chi_1$ can be either (semi-)prompt or long-lived. However, as the $\chi_i$ are neutral, unlike in the case of a long-lived $\phi$, the existence of a charged track cannot be exploited. At the same time, due to the small mass splitting among the $\chi_i$, the jets resulting from their decay tend to be very soft, rendering current displaced jet searches insensitive. However, the scenario can be constrained by $\slashed E_T$ searches, see Tab.~\ref{tab:lhcsearches}, mostly due to $\phi$-pair production and their  subsequent prompt decay into $\chi_2$ and/or $\chi_1$. We also note that the large lifetime of the neutral $\chi_i$ decaying into a quark pair makes it an interesting scenario for dedicated long-lived particle detectors outside the caverns of the main LHC experiments, such as MATHUSLA~\cite{Curtin:2018mvb}. 

Generally, the regions of uncovered parameter space in the conversion-driven freeze-out scenario feature a small mass splitting between the long-lived particle and dark matter, resulting in soft decay products that challenge current search strategies. Different classes of searches that cover the entire range of lifetimes and their application to conversion-driven DM freeze-out have recently been put forward in \cite{Heisig:2024xbh}.

\subsection{Model-specific signatures}

As discussed in Sec.~\ref{sec:SMSconst} and already found in Ref.~\cite{Acaroglu:2021qae}, same-sign $uu\to\phi\phi$ production significantly enhances the cross section for mediator pair production in our model. This process leads to much stronger exclusion limits on Majorana DM in final states where the mediator charge is not identified. In principle, these enhanced cross sections could be used to distinguish our Majorana flavoured DM model~\cite{Acaroglu:2021qae} from the corresponding flavoured DM model with Dirac DM, studied in Refs.~\cite{Blanke:2017tnb,Blanke:2020bsf}. In practice, however, a distinction based solely on cross sections is challenging. On the one hand, precise measurements of absolute cross sections are a difficult task at hadron colliders. On the other hand, the production of coloured particles at the LHC is typically subject to large higher-order QCD corrections which have not been calculated in our model. 
In what follows we hence investigate alternative options to discriminate between Dirac and Majorana flavoured DM, by making use of signatures that rely on both the  same-sign $uu\to\phi\phi$ production channel as well as the non-trivial flavour structure of our model.

\subsubsection{Same-sign di-top + \texorpdfstring{$\slashed E_T$}{missing transverse energy}}

As pointed out in Ref.~\cite{Acaroglu:2021qae} the process $uu\to\phi\phi$ with subsequent mediator decay to top and DM leads to the same-sign signature $tt+\slashed{E}_T$ with two positively charged top quarks, with cross sections in the fb regime. Experimentally, the distinction from the more common $t\bar t+\slashed E_T$ signature, prominent in supersymmetric models, is possible by requiring two positively charged leptons from semileptonic top decays. Such an analysis has for instance been performed in the CMS search for same-sign top signatures of Ref.~\cite{CMS:2020cpy}.

There, a supersymmetric model is considered in which a pair of gluinos is produced, both of which decay into a top or antitop associated with a stop. The stop is then assumed to decay into a light quark and a neutralino, giving rise to the same-sign signatures $ttjj + \slashed{E}_T$ and $\bar t\bar t jj+\slashed E_T$. However, since this search assumes a small mass splitting of roughly $20\,\mathrm{GeV}$ between the stop and the neutralino, the jets in the final state are very soft. Therefore, in order to obtain a rough estimate of the constraints such a search would impose on our model, we calculate the production rates of the two same-sign final states $tt+\slashed E_T$ as well as $\bar t\bar t+\slashed E_T$ and compare them with the upper limits provided in Ref.~\cite{CMS:2020cpy}. We also ignore the kinematic differences that arise from distinct spin-statistics, as $\phi$ is a scalar whereas gluinos are fermions. Given that we do not expect the efficiencies of the analysis to be strongly kinematic-dependent, we assume this to only cause a negligible difference in the upper limits from Ref.~\cite{CMS:2020cpy}.

For the calculation of the leading-order production cross section of the two same-sign final states we use a \textsc{FeynRules}~\cite{Alloul:2013bka} implementation of our model, generate a \texttt{UFO} file, and employ \textsc{MadGraph5\_aMC@NLO}~\cite{Alwall:2014hca} to generate events. We here consider the flavour- and CP-conserving case with $\theta_{ij} = \phi_{ij} = \delta_{ij} = \gamma_i = 0$, since we are interested in the strongest possible constraints such a search would place on the NP masses $m_\phi$ and $m_{\chi_3}$. Allowing for non-vanishing mixing angles would only reduce the branching ratio of a given flavour-conserving final state and thus reduce its signal cross section. For similar reasons we here also assume a degenerate mass spectrum $m_{\chi_i} = m_\chi$. In total we thus vary the three coupling parameters $D_{1,2,3}$ and the masses $m_\phi$ and $m_\chi$ independently. 

\begin{figure}
	\centering
		\begin{subfigure}[t]{0.49\textwidth}
		\includegraphics[width=\textwidth]{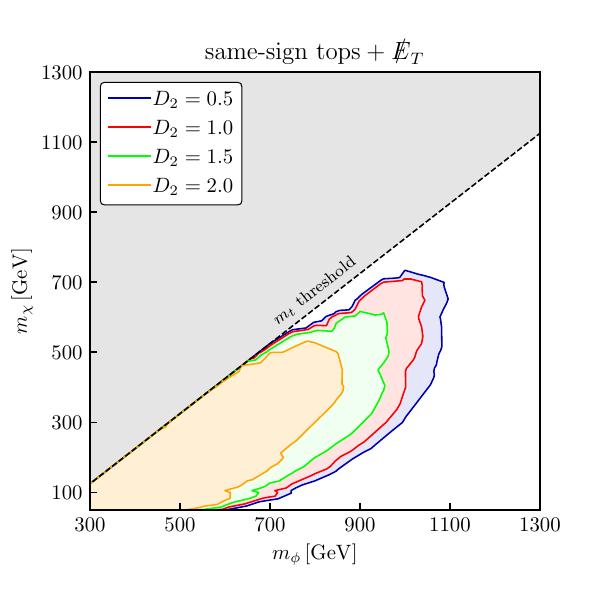}
		\caption{$D_1=D_3=2.0$}
		\end{subfigure}
		\begin{subfigure}[t]{0.49\textwidth}
		\includegraphics[width=\textwidth]{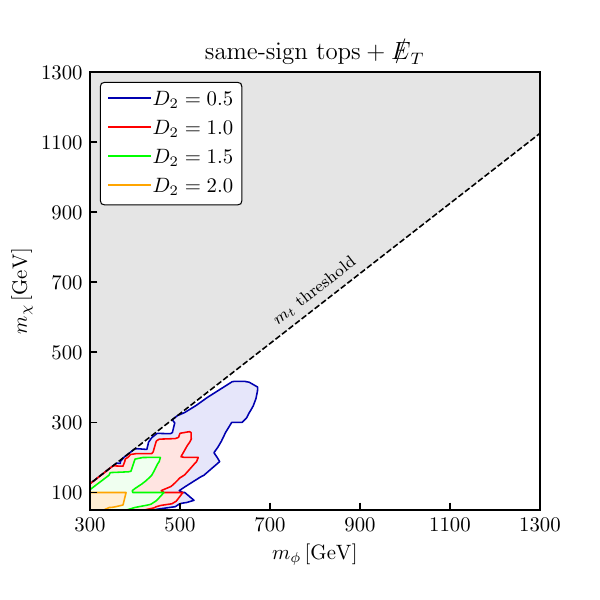}
		\caption{$D_1=1$, $D_3=2.0$}
		\end{subfigure}
	\caption{Estimated   constraints on the same-sign final states $tt+\slashed{E}_T$ and $\bar{t}\bar{t}+\slashed{E}_T$ obtained from Ref.~\cite{CMS:2020cpy}. The area under the curve is excluded.}
 \label{fig:collider:samesign}
	\end{figure}

The results of this procedure are summarised in Fig.~\ref{fig:collider:samesign}. In the left panel we show the resulting estimated reach
in the $m_\phi$--$m_\chi$ plane for $D_1=D_3=2.0$ and different values of $D_2$. Since a proton contains two valence up-quarks, aside from the enhancement from parton distribution functions, large DM--up couplings $D_1$ also enhance the production of a same-sign pair $\phi\phi$. On the other hand, large DM--top couplings $D_3$ are required in order to guarantee a sizeable branching ratio of the mediator into the relevant final states. The case considered in this panel hence yields the largest possible constraints, which for vanishing DM--charm couplings $D_2$ become maximal and reach up to $m_\phi \simeq 1100\,\mathrm{GeV}$. 

With respect to the DM mass $m_\chi$ we find that, in contrast to searches in opposite-sign final states, the largest exclusion is not obtained for vanishing masses $m_\chi \simeq 0$ but rather tends to grow with $m_\chi$. This is due to the fact that the $t$-channel exchange of two same-sign initial state quarks is governed by the Majorana mass term, which mixes the fields $\chi$ and $\chi^c$. So we see here that, depending on the mediator mass $m_\phi$, a same-sign top search could exclude DM masses up to $m_\chi \LessSim 750\,\mathrm{GeV}$. For increasing values of the DM--charm coupling $D_2$, we further find that the excluded range shrinks, as this at the same time reduces the branching ratio of the mediator into top-flavoured final states. For maximal couplings to all three generations, \emph{i.e.}~$D_i=2.0$, we find that the exclusions reach up to $m_\phi \LessSim 850\,\mathrm{GeV}$ and $m_\chi \LessSim 550\,\mathrm{GeV}$. 

For comparison, we also show the results for the same parameter choices but a smaller DM–up coupling $D_1 = 1.0$ in the right panel of Fig.~\ref{fig:collider:samesign}. While the results are qualitatively similar to the previous case, the same-sign $\phi$-pair production cross section is reduced accordingly, resulting in a much smaller exclusion region.

Comparing these limits with those found in Ref.~\cite{Acaroglu:2021qae}, we see that the exclusions obtained from the standard supersymmetric squark searches for jets+$\slashed E_T$ and tops+$\slashed E_T$ are significantly stronger than those from the same-sign $ttjj+\slashed E_T$ search considered here.  
The main origin of this difference is the requirement of two semileptonic tops in the final state, necessary in order to allow for charge identification. This suppresses the number of events down to about 4\% of the total $tt/\bar t\bar t+\slashed E_T$ rate, and is intrinsic to the determination of the top quark charge. In the next section we therefore explore how to test the Majorana nature of flavoured DM by making use of the top quark charge in single-top events.

\subsubsection{Single-top charge asymmetry}

Mediator pair production with subsequent decays into $j+\slashed E_T$ and $t+\slashed E_T$, respectively, induces the flavour-violating final state $tj+\slashed E_T$. For the case of flavoured Dirac DM the prospects of a dedicated search have been analysed in Ref.\cite{Blanke:2020bsf}, with the result that the region of testable parameter space can be significantly extended relative to only the flavour-conserving $jj+\slashed E_T$ and $t\bar t+\slashed E_T$. Noteworthy, the analysis strategy proposed in Ref.\cite{Blanke:2020bsf} involves a semileptonic top quark in the final state, opening up the possibility to straightforwardly extend it with charge identification.

In the case of Dirac DM only opposite-sign $\phi^\dagger\phi$ pairs are produced, and therefore the cross section of final states with a top or anti-top are equal: 
\begin{equation}
    \sigma_\text{Dirac}(tj + \slashed E_T) = \sigma_\text{Dirac}(\bar t j + \slashed E_T) \,.
\end{equation}
For Majorana DM however, also $\phi\phi$ and $\phi^\dagger\phi^\dagger$ pairs are produced, with the former being substantially enhanced by the valence up-quark content of the proton. Hence we predict
\begin{equation}
    \sigma_\text{Majorana}(tj + \slashed E_T) > \sigma_\text{Majorana}(\bar t j + \slashed E_T) \,,
\end{equation}
where the magnitude of the difference depends on the size of the Majorana DM mass as well as its coupling strength to up-quarks.
As a quantitative measure of this effect, we introduce the charge asymmetry
\begin{equation}
    a_{tj} = \frac{\sigma(tj+\slashed E_T)-\sigma(\bar tj+ \slashed E_T)}{\sigma(tj+\slashed E_T)+\sigma(\bar tj+\slashed E_T)}\,,
\end{equation}
Following the arguments given above, we expect\footnote{In the Dirac case, processes with mediator single-production could give a non-zero contribution to $a_{tj}$, however these effects are generally suppressed relative to the QCD-induced mediator pair-production and become important only in the limit of large mediator masses.}
\begin{eqnarray}
    \text{Dirac DM} & \Rightarrow & a_{tj} \simeq 0\,, \\
     \text{Majorana DM} & \Rightarrow & a_{tj} > 0 \,.
\end{eqnarray}

\begin{figure}[t!]
	\centering
	\begin{subfigure}[t]{0.49\textwidth}
	\includegraphics[width=\textwidth]{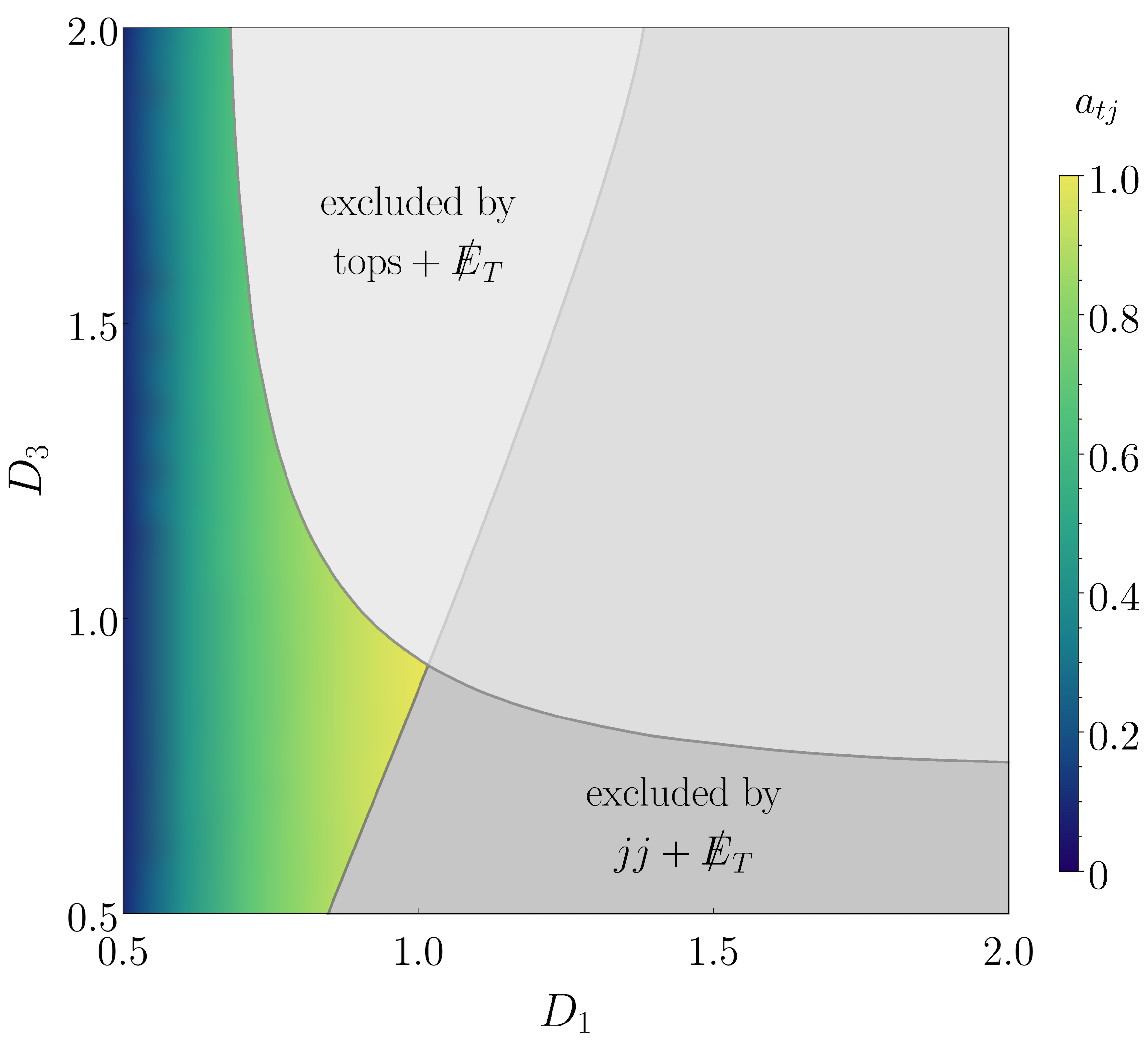}
	\end{subfigure}
    \begin{subfigure}[t]{0.49\textwidth}
	\includegraphics[width=0.90\textwidth]{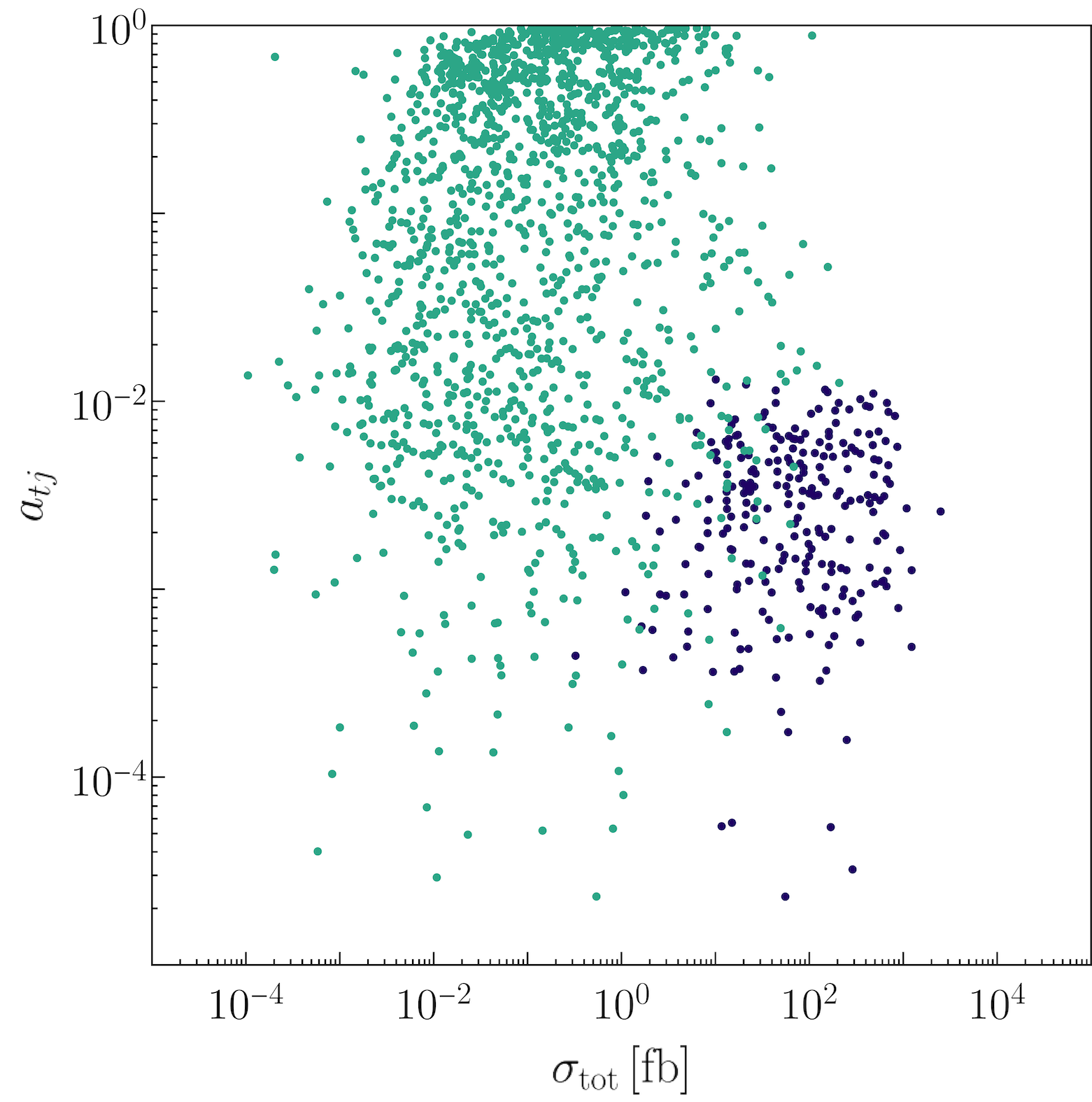}
	\end{subfigure}
 \caption{Left panel: Prediction of the charge asymmetry, $a_{tj}$, (colour code) at the 14\,TeV LHC in comparison with constraints from tops+$\slashed E_T$ (light grey) and jets+$\slashed E_T$ (darker grey)  for $D_2=0$, $m_\phi = 1200\,\mathrm{GeV}$ and $m_\chi = 400\,\mathrm{GeV}$. Right panel: Prediction of $a_{tj}$ based on the viable parameter space for the canonical freeze-out scenario (green) and the C$_\chi 1_u$ scenario (blue).  
 }
 \label{fig:asym}
	\end{figure}

To estimate the feasibility of measuring $a_{tj}$ at future LHC runs, we determine the $tj+\slashed E_T$ and $\bar tj+\slashed E_T$ cross sections using \textsc{MadGraph5\_aMC@NLO}~\cite{Alwall:2014hca},  from which we calculate $a_{tj}$. We consider a center-of-mass energy of $\sqrt{s}=14\,$TeV\@. The result is shown in Fig.~\ref{fig:asym}. In the left panel we present the prediction for the charge asymmetry in the coupling plane $(D_1, D_3)$ for a benchmark scenario of $m_\phi = 1200\,\mathrm{GeV}$, $m_{\chi_i} = 400\,\mathrm{GeV}$ and $D_2=0$. Parameter regions excluded by jets+$\slashed E_T$ and tops+$\slashed E_T$ according to the analysis in Ref.\ \cite{Acaroglu:2021qae} are shaded in grey. The asymmetry is found independent of the value of $D_3$ but grows with increasing $D_1$, as expected from the underlying process of same-sign $\phi$ pair-production. Large values $a_{tj}\simeq 1$ can be reached close to the excluded region, suggesting relevant LHC production cross sections. 

To better assess the possibility of measuring $a_{tj}$ at the LHC, the right panel of Fig.\ \ref{fig:asym} relates the charge asymmetry to the total cross section $\sigma_\text{tot}=\sigma(tj+\slashed E_T)+\sigma(\bar tj+\slashed E_T)$, displaying viable points identified in our numerical scans. In the canonical freeze-out scenario (green points) we find total cross sections  up to $\sigma_\text{tot}\simeq 100\,\text{fb}$ for $\mathcal{O}(1)$ values of $a_{tj}$, indicating that the charge asymmetry is indeed a promising observable to test the nature of DM\@. For the conversion-driven freeze-out scenario C$_\chi 1_u$, on the other hand, we find that the asymmetry is  very small, $a_{tj}<10^{-2}$. This is because in this scenario the up-quark coupling $D_1$ is strongly suppressed by construction, so that the same-sign $\phi$ pair-production process becomes irrelevant. In the C$_\phi$ freeze-out scenarios, final states with top quarks are kinematically forbidden. The single-top charge asymmetry thus not only tests the nature of DM, but its measurement different from zero would also allow to strongly constrain the conversion-driven freeze-out scenarios studied above.

\section{Conclusions}
\label{sec:Conclude}

In this paper, we study the phenomenology of a simplified Dark Matter (DM) model within the Dark Minimal Flavour Violation (DMFV) framework, where the SM is supplemented by a Majorana fermion flavour triplet, which provides a DM candidate, and a coloured scalar, which acts as a $t$-channel mediator for DM interactions with the right-handed up-type quarks. 

In the first part of our analysis, we explored the allowed parameter space of the model, considering constraints from flavour, direct and indirect detection observables, and in particular the relic density measurement, by performing a comprehensive set of scans of the 18 model parameters. We found realisations of a wide range of DM freeze-out scenarios that satisfy all constraints. 

First, coannihilation effects with both the heavier DM triplet states or the coloured scalar mediator open up a cosmologically viable parameter space for mass splittings of $\lesssim10\%$ that has not been discussed in previous studies of the model. Second, and more importantly, we found realisations of conversion-driven freeze-out in the model. This scenario is characterised by a very weak DM coupling that renders conversion processes between DM and coannihilator semi-efficient, which in turn initiates freeze-out. In order to support sufficient dilution of dark sector particles before freeze-out, the annihilation cross section of the coannihilator must be sufficiently large. This requirement is met for the case where the coloured mediator is the dominant coannihilator (scenarios C$_\phi1_u$ and C$_\phi2_u$). These scenarios populate the parameter region for very small mass splittings between DM and $\phi$ of less than a few percent, which is excluded for canonical freeze-out. We also found realisations where $\chi_1$ (the heaviest Majorana triplet state) acts as the dominant coannihilator. This case (C$_\chi1_u$) requires a significant coupling of $\chi_1$ to the top quark, since the $\chi_1$ pair annihilation cross section is helicity suppressed, and hence a large hierarchy within the coupling matrix. 

Using the program package \textsc{SModelS}, we have confronted the allowed parameter space with a large number of results from new physics searches at the LHC\@. In the canonical scenario, exclusions arise from searches for jets+$\slashed E_T$ or tops+$\slashed E_T$, which cover a large range in the considered masses of new particles. The model exhibits a much richer phenomenological structure than a typical simplified model with a DM singlet state, as used in the interpretation of the LHC results by the collaborations. In our model, on the one hand, same-sign mediator production via the $t$-channel $\chi_i$ diagram can significantly enhance the cross section with respect to gluon-initiated $\phi$ production, leading to the exclusion of large new-physics particle masses. On the other hand, metastable intermediate $\chi$ states arising from small mass splittings between them can make the application of $\slashed E_T$ searches questionable and call for closer investigation.

In the conversion-driven freeze-out scenarios, long-lived particle searches provide a potentially sensitive probe of our model. However, current searches for displaced vertices target hard displaced jets and thus lack sensitivity to these scenarios, which involve soft jets due to small mass splittings. Accordingly, the charged track of R-hadrons formed by pair-produced metastable mediator particles leads to the only signature well covered by current long-lived particle searches, excluding DM masses up to about 1.3\,TeV. On the other hand, parts of the conversion-driven freeze-out region can be probed by conventional $\slashed E_T$ searches, in particular in the C$_\phi2_u$ scenario.

Finally, we explored two model-specific signatures that could shed light on the Majorana nature of the DM multiplet -- a same-sign di-top+$\slashed E_T$ signature and a single-top charge asymmetry.  To assess the discovery potential of the former, we performed a naive recast of an existing CMS search \cite{CMS:2020cpy} for $ttjj+\slashed E_T$ final states and found that the resulting limits would not be 
competitive with the standard jets$+\slashed E_T$ and tops$+\slashed E_T$ analyses, mainly
due to the requirement that both tops decay semileptonically. We thus introduced the single-top charge asymmetry $a_{tj}$, for which a non-zero value would again indicate the Majorana nature of DM. Numerically, we found parts of the parameter space where both the asymmetry and the underlying single-top total cross section are large, encouraging further studies at the LHC\@.

\section*{Acknowledgements} 

We thank Margarete M\"uhlleitner for her collaboration in the early stages of this work. 
This research was supported by the Deutsche Forschungsgemeinschaft (DFG, German Research Foundation) under grant 396021762 -- TRR 257.  J.H.~acknowledges support by the Alexander von Humboldt foundation via the Feodor Lynen Research Fellowship for Experienced Researchers and the Feodor Lynen Return Fellowship.

\appendix

\section{Cross sections and rates}\label{app:XS}

In this appendix, we provide analytic expressions for the relevant annihilation and conversion processes used in our analysis. The thermal average for the $2\to2$ cross sections is obtained by~\cite{Edsjo:1997bg}
\begin{equation}
\langle  \sigma v \rangle_{ij}=   \sum_{kl} \frac{g_i g_j\,T}{8\pi^4 \,n_i^{\eq}n_j^{\eq}}\int  \D s \,\sqrt{s}\, \,p_{\text{COM}}(s,m_i,m_j)^2  \, \sigma_{ij\to kl}(s) \,K_1\!\left(\frac{\sqrt{s}}{T}\right) \,,
\label{eq:thermalav}
\end{equation}  
where
\begin{equation}
p_{\text{COM}}(s,m_i,m_j)= \frac{\sqrt{s- (m_i+m_j)^2}\sqrt{s- (m_i-m_j)^2}}{2\sqrt{s}}\,.
\end{equation}
Note that in this appendix, $s$ and $t$ denote the respective Mandelstam variables. The former shall not to be confused with the entropy density denoted by $s$ in the main text. 

For all cross sections the lower and upper integration limits of the Mandelstam variable $t$, $t_{\mathrm{min}}$ and $t_{\mathrm{max}}$, respectively, are given by:
\begin{align} 
    t_{\mathrm{min}}(s, m_1, m_2, m_3, m_4) = & \,m_1^2 + m_3^2 - 2\sqrt{p_{\mathrm{COM}}(s, m_1, m_2)^2 + m_1^2} \sqrt{p_{\mathrm{COM}}(s, m_3, m_4)^2 + m_3^2}\nonumber \\
    & - 2 p_{\mathrm{COM}}(s, m_1, m_2) \,p_{\mathrm{COM}}(s, m_3, m_4)\\
    t_{\mathrm{max}}(s, m_1, m_2, m_3, m_4) = & \,m_1^2 + m_3^2 - 2\sqrt{p_{\mathrm{COM}}(s, m_1, m_2)^2 + m_1^2} \sqrt{p_{\mathrm{COM}}(s, m_3, m_4)^2 + m_3^2} \nonumber\\
    & + 2 p_{\mathrm{COM}}(s, m_1, m_2) \,p_{\mathrm{COM}}(s, m_3, m_4)
\end{align}

\subsection{Cross sections for annihilation processes}

\noindent $\bullet\;\phi \, \phi^\dagger \rightarrow g  g$: 
\begin{align} 
    \begin{split}
    \sigma_{\phi \phi^\dagger \rightarrow g g} (s) = & \; \frac{g_s^4}{54 \pi s^2(4 m_\phi^2 - s)} \left[\left(62 m_\phi^2 + 5 s\right) \sqrt{s(s-4 m_\phi^2)} \rule{0em}{8mm}\right.\\
        &  \left. \rule{0em}{8mm}+ 8\left(4m_\phi^2 s + m_\phi^4\right) ~ \mathrm{log}\!\left( \frac{\sqrt{s(s-4 m_\phi^2)} - s}{\sqrt{s(s-4 m_\phi^2)} + s}\right)\, \right]
    \end{split}
\end{align}

\noindent $\bullet\;\phi \, \phi^\dagger \rightarrow q_i \bar{q_i}$ for up-type quarks:  
\begin{align} 
    \begin{split}
    \sigma_{\phi \phi^\dagger \rightarrow  q_i \bar{q_i}} (s) = & \,\frac{1}{144 \pi  s^3 (s-4 m_{\phi }^2)} \int_{t_{\mathrm{min}}(s)}^{t_{\mathrm{max}}(s)} \!\D t \Bigg( 16 s  \left(m_q^4+2 t m_{\phi }^2-m_{\phi }^4-t (s+t)\right) \sum_j \frac{ g_s^2 \lvert\lambda _{ij}\rvert^2}{(t-m_{\chi_j}^2)} \\
    & - 16 g_s^4 \left(-m_q^2 (-2 m_{\phi }^2+s+2 t)+m_q^4-2 t m_{\phi }^2+m_{\phi }^4+t (s+t)\right)\\
    & - 9 s^2 \left(-2 m_q^2 m_{\phi }^2+m_q^4-2 t m_{\phi }^2+m_{\phi }^4+t (s+t)\right) \\ 
    & \hspace{-8ex}\times
    \frac{\left(\lvert\lambda _{i1}\rvert^2 (t-m_{\chi_2}^2) (t-m_{\chi_3}^2)+ \lvert\lambda _{i2}\rvert^2 (t-m_{\chi_1}^2) (t-m_{\chi_3}^2)+\lvert\lambda _{i3}\rvert^2 (t-m_{\chi_1}^2) (t-m_{\chi_2}^2)\right)^2}{(t-m_{\chi_1}^2) (t-m_{\chi_2}^2) (t-m_{\chi_3}^2)} \Bigg), 
    \end{split}
\end{align}
where $t_{\mathrm{min}/\mathrm{max}}(s) = t_{\mathrm{min}/\mathrm{max}}(s, m_{\phi}, m_{\phi}, m_{q_i}, m_{q_i})$.

\bigskip

\noindent $\bullet\;\phi \, \phi^\dagger \rightarrow q_i  \bar{q_i}$ for down-type quarks: 
\begin{equation}
    \sigma_{\phi \phi^\dagger \rightarrow  q_i \bar{q_i}} (s) = -\int_{t_{\mathrm{min}}(s)}^{t_{\mathrm{max}}(s)} \D t \,\frac{g_s^4 \left[m_q^2 \left(2 m_{\phi }^2-s-2 t\right)+m_q^4-2 t m_{\phi }^2+m_{\phi }^4+t (s+t)\right]}{9 \pi s^3 \left(s-4 m_{\phi }^2\right)},
\end{equation}
where $t_{\mathrm{min}/\mathrm{max}}(s) = t_{\mathrm{min}/\mathrm{max}}(s, m_{\phi}, m_{\phi}, m_{q_i}, m_{q_i})$.

\bigskip

\noindent $\bullet\;\phi \, \phi^\dagger \rightarrow q_i  \bar{q_j}$, $i\neq j$: 
\begin{align}
    \begin{split}
       \sigma_{\phi \phi^\dagger \rightarrow  q_i \bar{q_j}} (s) = & -\int_{t_{\mathrm{min}}(s)}^{t_{\mathrm{max}}(s)} \D t \left(-2 m_{\phi }^2 \left(m_{q_i}^2+t\right)+t m_{q_i}^2+m_{q_i}^4-t m_{q_j}^2+m_{\phi }^4+t (s+t)\right) \\
       & \hspace{-8ex}\times \frac{\lvert \lambda _{i1} \lambda _{j1} (t-m_{\chi_2}^2) (t-m_{\chi_3}^2)+ \lambda _{i2} \lambda _{j2} (t-m_{\chi_1}^2) (t-m_{\chi_3}^2)+\lambda _{i3} \lambda _{j3} (t-m_{\chi_1}^2) (t-m_{\chi_2}^2)\rvert^2}{16 \pi s (s-4 m_{\phi }^2) (t-m_{\chi_1}^2)^2 (t-m_{\chi_2}^2)^2 (t-m_{\chi_3}^2)^2} ,
    \end{split}
\end{align}
where $t_{\mathrm{min}/\mathrm{max}}(s) = t_{\mathrm{min}/\mathrm{max}}(s, m_{\phi}, m_{\phi}, m_{q_i}, m_{q_j})$.

\bigskip
\noindent $\bullet\;\phi \, \phi \rightarrow q_i  q_j$, $i\neq j$: 
\begin{align}
    \begin{split}
     \sigma_{\phi \phi \rightarrow  q_i q_j} (s) = &\,  \sum_{k,l} \frac{\lambda _{ik}^* \lambda _{il} \lambda _{jk}^* \lambda _{jl} m_{\chi_k} m_{\chi_l} \left(-m_{q_i}^2-m_{q_j}^2+s\right)}{48 \pi s \left(s-4 m_{\phi }^2\right)} \int_{t_{\mathrm{min}}(s)}^{t_{\mathrm{max}}(s)} \D t\, \frac{1}{(t-m_{\chi_k}^2) (t-m_{\chi_l}^2)}\\
    &  \times \Bigg(\frac{3 (m_{q_i}^2 + m_{q_j}^2)^2 -(2 m_{q_i}^2 +2 m_{q_j}^2) \big(2 m_{\chi_k}^2+2 m_{\chi_l}^2-6 m_{\phi }^2+3 s+2 t\big)}{ (-m_{q_i}^2-m_{q_j}^2+m_{\chi_k}^2-2 m_{\phi }^2+s+t) (-m_{q_i}^2-m_{q_j}^2+m_{\chi_l}^2-2 m_{\phi }^2+s+t)} \\
    &  \hspace{-12.5ex} +\frac{4 \big[-m_{\phi }^2 \big(2 m_{\chi_k}^2+2 m_{\chi_l}^2+3 s+2 t\big)+m_{\chi_l}^2 \left(2 m_{\chi_k}^2+s\right)+s m_{\chi_k}^2+3 m_{\phi }^4\big]+3 s^2+4 s t+4 t^2}{(-m_{q_i}^2-m_{q_j}^2+m_{\chi_k}^2-2 m_{\phi }^2+s+t) (-m_{q_i}^2-m_{q_j}^2+m_{\chi_l}^2-2 m_{\phi }^2+s+t)}\Bigg),
    \end{split}
\end{align}
where $t_{\mathrm{min}/\mathrm{max}}(s) = t_{\mathrm{min}/\mathrm{max}}(s, m_{\phi}, m_{\phi}, m_{q_i}, m_{q_j})$.

\bigskip
\noindent $\bullet\;\phi \, \phi \rightarrow q_i  q_i$:
\begin{equation}
    \sigma_{\phi \phi \rightarrow  q_i q_i} (s) = \frac{1}{2} \sigma_{\phi \phi \rightarrow  q_i q_j} (s)\big |_{ i=j}
\end{equation}

\bigskip
\noindent $\bullet\;\phi \, \chi_i \rightarrow q_j  g$: 
\begin{align}
    \begin{split}
       \sigma_{\phi \chi_i \rightarrow  q_j g} (s) = & \,\frac{g_s^2 \lvert\lambda _{ji}\rvert^2}{48  \pi s \big(s-m_{q_j}^2\big)^2\, p_{\mathrm{COM}}(s, m_\phi, m_{\chi_i})^2} \\ 
       &\times \int_{t_{\mathrm{min}}(s)}^{t_{\mathrm{max}}(s)} \D t \, \Big( m_{q_j}^2 \Big\{-t \left(-4 s m_{\chi_i}^2+2 m_{\chi_i}^4+2 s^2+4 s t+t^2\right)   \\
       &  \;-m_{\phi }^4 \left(4 m_{\chi_i}^2+4 s+5 t\right)+m_{\phi }^2 \left[4 t m_{\chi_i}^2+2 m_{\chi_i}^4+t (4 s+3 t)\right]+3 m_{\phi }^6\Big\} \\
       & \;+m_{q_j}^4 \Big\{t \left(-2 m_{\chi_i}^2+2 s+t\right)+2 s m_{\phi }^2+m_{\phi }^4\Big\}-2 m_{q_j}^6 m_{\phi }^2 \\
       &  \;+s \Big\{\left(m_{\phi }^4+t^2\right) \left(-m_{\phi }^2+s+t\right)+2 m_{\chi_i}^2 \left(m_{\phi }^4-t (s+t)\right)+2 m_{\chi_i}^4 \left(t-m_{\phi }^2\right)\Big\}\Big),
    \end{split}
\end{align}
where $t_{\mathrm{min}/\mathrm{max}}(s) = t_{\mathrm{min}/\mathrm{max}}(s, m_{\chi_i}, m_{\phi}, m_{q_j}, 0)$.

\bigskip
\noindent $\bullet\;\chi_i \, \chi_j \rightarrow q_k  q_l$:
\begin{align}
    \begin{split}
    \sigma_{\chi_i \chi_j \rightarrow  q_k q_l} (s) = & \;\frac{3}{256 \pi s\, p_{\mathrm{COM}}(s, m_{\chi_i}, m_{\chi_j})^2 } \\
    & \times \int_{t_{\mathrm{min}}(s)}^{t_{\mathrm{max}}(s)} \D t\, \Bigg(\!-\frac{2 \mathrm{Re}(\lambda _{ki} \lambda _{kj} \lambda _{li} \lambda _{lj}) \, m_{\chi_i} m_{\chi_j} \big(s -m_{q_k}^2-m_{q_l}^2\big)}{\big(t-m_{\phi }^2\big) \big(m_{q_k}^2+m_{q_l}^2+m_{\chi_i}^2+m_{\chi_j}^2-m_{\phi }^2-s-t\big)}\\
    & +\frac{\lvert \lambda _{ki}\rvert^2 \lvert\lambda _{lj}\rvert^2 \big(-m_{q_k}^2-m_{\chi_i}^2+t\big) \big(-m_{q_l}^2-m_{\chi_j}^2+t\big)}{\big(t-m_{\phi }^2\big)^2}\\
    &  +\frac{\lvert\lambda _{kj}\rvert^2 \lvert\lambda _{li}\rvert^2 \big(-m_{q_k}^2-m_{\chi_j}^2+s+t\big) \big(-m_{q_l}^2-m_{\chi_i}^2+s+t\big)}{\big(m_{q_k}^2+m_{q_l}^2+m_{\chi_i}^2+m_{\chi_j}^2-m_{\phi }^2-s-t\big){}^2}\Bigg),
    \end{split}
\end{align}
where $t_{\mathrm{min}/\mathrm{max}}(s) = t_{\mathrm{min}/\mathrm{max}}(s, m_{\chi_i}, m_{\chi_j}, m_{q_k}, m_{q_l})$.

\subsection{Cross sections for conversion processes (scatterings)}

For the conversion via scattering we take into account the leading diagrams in $\alpha_\text{s}$  neglecting sub-leading electroweak contributions.
\vspace{0.8ex}

\noindent $\bullet\;\phi , q_j \rightarrow \chi_i  g$: 
\begin{align}
    \begin{split}
    \sigma_{\phi  q_j \rightarrow \chi_i g} (s) = & \frac{g_s^2 \lvert \lambda _{ji}\rvert^2}{72 \pi s\, p_{\mathrm{COM}}(s, m_{\phi}, m_{q_j})^2 } \int_{t_{\mathrm{min}}(s)}^{t_{\mathrm{max}}(s)} \D t\, \Bigg( \frac{2 m_{q_j}^4 \big( s-m_{\chi_i}^2\big) }{\big(t-m_{\phi }^2\big) \big(-m_{\chi_i}^2-m_{\phi }^2+s+t\big)^2} \\
    & + \frac{2 m_{q_j}^2 \big\{2 m_{\chi_i}^2 \big(t m_{\phi }^2-m_{\phi }^4+s t\big)+s \big[m_{\phi }^4-t (s+t)\big]-t m_{\chi_i}^4\big\}}{\big(t-m_{\phi }^2\big)^2 \big(-m_{\chi_i}^2-m_{\phi }^2+s+t\big)^2} \\
    &  + \frac{t \big[s t-m_{\chi_i}^2 (2 s+t)\big]+m_{\phi }^4 \big(s-3 m_{\chi_i}^2\big)+2 m_{\chi_i}^2 m_{\phi }^2 \big(m_{\chi_i}^2+t\big)}{\big(t-m_{\phi }^2\big)^2 \big(-m_{\chi_i}^2-m_{\phi }^2+s+t\big)} \Bigg),
    \end{split}
\end{align}
where $t_{\mathrm{min}/\mathrm{max}}(s) = t_{\mathrm{min}/\mathrm{max}}(s, m_{q_j}, m_{\phi}, m_{\chi_i}, 0)$.

\bigskip
\noindent $\bullet\;\phi \, g \rightarrow \chi_i  q_j$: 
\begin{align}
    \begin{split}
     \sigma_{\phi g \rightarrow \chi_i q_j} (s) = & \;\frac{g_s^2 \lvert\lambda _{ji}\rvert^2}{192 \pi s (s-m_{\phi }^2)^2\, p_{\mathrm{COM}}(s, m_{\phi }, 0)^2 } \\
     &\times \int_{t_{\mathrm{min}}(s)}^{t_{\mathrm{max}}(s)} \D t \,\Bigg(-\frac{m_{q_j}^4 \big[s \left(-2 m_{\chi_i}^2+s+2 t\right)+2 t m_{\phi }^2+m_{\phi }^4\big]}{(t-m_{q_j}^2)^2} \\
     & \; + \frac{m_{q_j}^2 \big[ s \left(2 m_{\chi_i}^4-4 t m_{\chi_i}^2+s^2+4 s t+2 t^2\right)-m_{\phi }^2 \left(4 s m_{\chi_i}^2+2 m_{\chi_i}^4+3 s^2+4 s t\right)\big]}{(t-m_{q_j}^2)^2} \\
     & + \; \frac{m_{q_j}^2 \big[m_{\phi }^4 \left(4 m_{\chi_i}^2+5 s+4 t\right)-3 m_{\phi }^6\big]+2 m_{q_j}^6 m_{\phi}^2}{(t-m_{q_j}^2)^2} \\
     & + \; \frac{t \big\{\big(m_{\phi }^4+s^2\big) \big(m_{\phi }^2-s-t\big)+2 m_{\chi_i}^2 \big[s (s+t)-m_{\phi }^4\big]+2 m_{\chi_i}^4 \big(m_{\phi }^2-s\big)\big\}}{(t-m_{q_j}^2)^2}\Bigg),
    \end{split}
\end{align}
where $t_{\mathrm{min}/\mathrm{max}}(s) = t_{\mathrm{min}/\mathrm{max}}(s, m_{\phi}, 0, m_{\chi_i}, m_{q_j})$. Note that we regularised the soft divergence in the cross section for $\phi \, g \rightarrow \chi_i  q_j$ by introducing a thermal mass for the gluon.

\bigskip
\noindent $\bullet\;\chi_i \,q_k \rightarrow \chi_j  q_l$:
\begin{align}
    \begin{split}
    \sigma_{\chi_i q_k \rightarrow \chi_j q_l} (s) = & \;\frac{1}{128 \pi s\, p_{\mathrm{COM}}(s, m_{\chi_i}, m_{q_k})^2 } \\
    &\times \int_{t_{\mathrm{min}}(s)}^{t_{\mathrm{max}}(s)} \D t \,\Bigg( \frac{\lvert\lambda_{ki}\rvert^2 \lvert\lambda_{lj}\rvert^2 \big(-m_{q_k}^2-m_{\chi_i}^2+s\big) \big(-m_{q_l}^2-m_{\chi_j}^2+s\big)}{\big(\Gamma_\phi^2-2 s\big) m_{\phi }^2+m_{\phi }^4+s^2} \\
    & \; + \frac{\lvert\lambda_{kj}\rvert^2 \lvert\lambda_{li}\rvert^2 \big(-m_{q_k}^2-m_{\chi_j}^2+t\big) \big(-m_{q_l}^2-m_{\chi_i}^2+t\big)}{\big(\Gamma_\phi^2-2 t\big) m_{\phi }^2+m_{\phi }^4+t^2} \\
    & \hspace{-5ex}+ \frac{2 \mathrm{Re}(\lambda_{ki} \lambda_{kj} \lambda_{li} \lambda_{lj})\,  m_{\chi_i} m_{\chi_j} \big[m_{\phi}^2 \Gamma_\phi^2+ (m_{\phi}^2-s)(m_{\phi}^2-t)\big] \big(-m_{\chi_i}^2-m_{\chi_j}^2+s+t\big)}{\big[\big(\Gamma_\phi^2-2 s\big) m_{\phi }^2+m_{\phi }^4+s^2\big] \big[\big(\Gamma_\phi^2-2 t\big) m_{\phi }^2+m_{\phi }^4+t^2\big]} \Bigg),
    \end{split}
\end{align}
where $t_{\mathrm{min}/\mathrm{max}}(s) = t_{\mathrm{min}/\mathrm{max}}(s, m_{\chi_i}, m_{q_k}, m_{q_l}, m_{\chi_j})$.

\subsection{Decay rates}

\noindent $\bullet\;\phi \rightarrow \chi_i  q_j$:
\begin{equation}
    \Gamma_{\phi \rightarrow \chi_i q_j} = \frac{|\lambda_{ji}|^2}{8\pi m_{\phi}^2} (m_{\phi}^2-m_{q_i}^2-m_{\chi_j}^2) \,p_{\mathrm{COM}}(m_{\phi}^2, m_{q_i}, m_{\chi_j})
\end{equation}

\bigskip
\noindent $\bullet\;\chi_i \rightarrow \chi_j  q_k  q_l$: 
\begin{align}
    \begin{split}
    \Gamma_{\chi_i \rightarrow \chi_j q_k q_l} = & \frac{3}{128 \pi m_{\chi_i}^3} \,\int_{m_{12}^{\mathrm{min}}}^{m_{12}^{\mathrm{max}}} \D m_{12} \int_{m_{23}^{\mathrm{min}}(m_{12})}^{m_{23}^{\mathrm{max}}(m_{12})} \D m_{23} \\
    & \times \Bigg(\frac{2 \mathrm{Re}(\lambda_{ki} \lambda_{kj} \lambda_{li} \lambda_{lj})\, m_{\chi_i} m_{\chi_j} \big(m_{q_k}^2+m_{q_l}^2-m_{12}\big)}{\big(m_{23}-m_{\phi }^2\big) \big(-m_{q_k}^2-m_{q_l}^2-m_{\chi_i}^2-m_{\chi_j}^2+m_{\phi }^2+m_{12}+m_{23}\big)}
    \\
    & \;-\frac{\lvert\lambda_{ki}\rvert^2 \lvert\lambda_{lj}\rvert^2 \big(-m_{q_k}^2-m_{\chi_i}^2+m_{23}\big) \big(-m_{q_l}^2-m_{\chi_j}^2+m_{23}\big)}{\big(m_{23}-m_{\phi }^2\big){}^2} \\
    & \; +\frac{\lvert\lambda_{kj}\rvert^2 \lvert\lambda_{li}\rvert^2 \big(-m_{q_k}^2-m_{\chi_j}^2+m_{12}+m_{23}\big) \big(m_{q_l}^2+m_{\chi_i}^2-m_{12}-m_{23}\big)}{\big(m_{q_k}^2+m_{q_l}^2+m_{\chi_i}^2+m_{\chi_j}^2-m_{\phi }^2-m_{12}-m_{23}\big){}^2}\Bigg),
    \end{split}
\end{align}
where $m_{12}^{\mathrm{min}} = (m_{q_k} + m_{q_l})^2$,
$m_{12}^{\mathrm{max}} = (m_{\chi_i} - m_{\chi_j})^2$, and
\begin{align}
m_{23}^{\mathrm{min}}(m_{12}) = &\;\frac{(m_{q_l}^2 -m_{q_k}^2 + m_{\chi_i}^2 - m_{\chi_j}^2)^2}{4m_{12}}\nonumber 
\\
&- \left(\sqrt{\frac{(m_{12} - m_{q_k}^2 + m_{q_l}^2)^2}{4m_{12}} - m_{q_l}^2} + \sqrt{\frac{(m_{\chi_i}^2 - m_{12} - m_{\chi_j}^2)^2}{4m_{12}} - m_{\chi_j}^2}\,\right)^{\!2},
\\
m_{23}^{\mathrm{max}}(m_{12}) = &\;\frac{(m_{q_l}^2 -m_{q_k}^2+ m_{\chi_i}^2 - m_{\chi_j}^2)^2}{4m_{12}} \nonumber
\\
& + \left(\sqrt{\frac{(m_{12} - m_{q_k}^2 + m_{q_l}^2)^2}{4m_{12}} - m_{q_l}^2} + \sqrt{\frac{(m_{\chi_i}^2 - m_{12} - m_{\chi_j}^2)^2}{4m_{12}} - m_{\chi_j}^2}\,\right)^{\!2}.
\end{align}

\section{Evolution of freeze-out abundances}
\label{qpp:abund}

\begin{figure}[t]
	\centering
    \begin{subfigure}[t]{0.49\textwidth}
	\includegraphics[width=\textwidth]{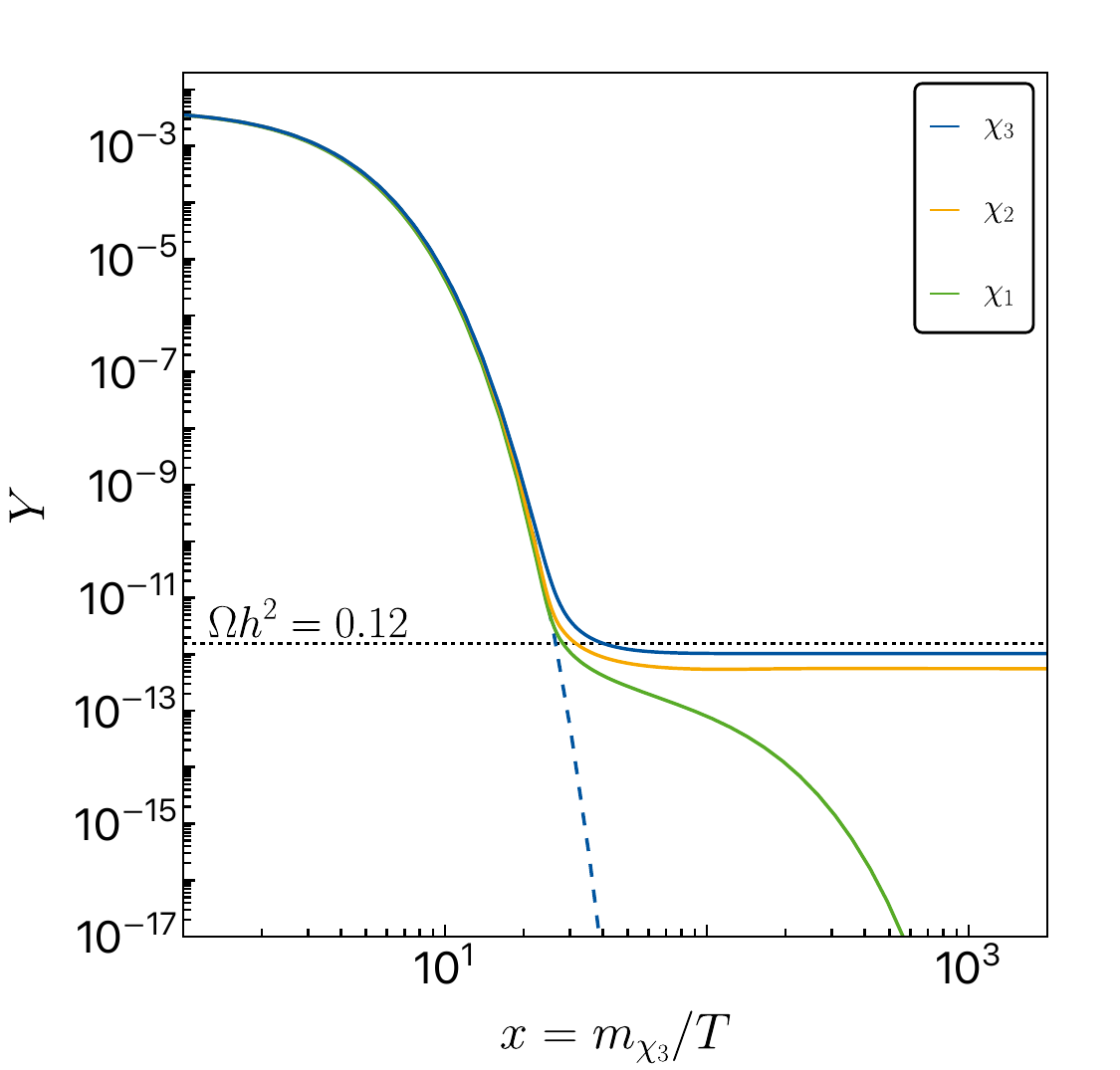}
	\end{subfigure}
	\begin{subfigure}[t]{0.49\textwidth}
	\includegraphics[width=\textwidth]{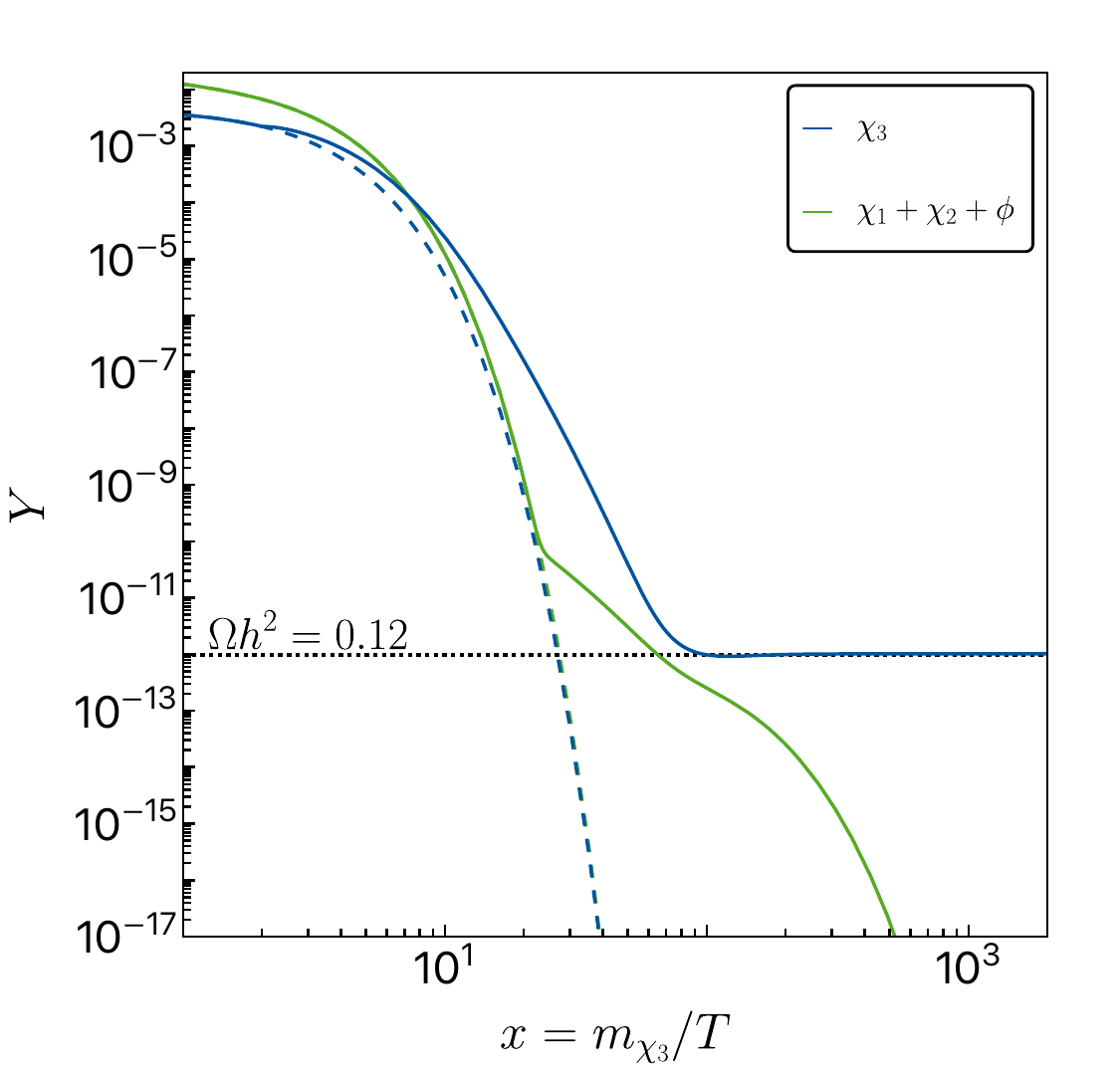}
	\end{subfigure}
	\caption{Example of the evolution of the abundances of different particles in two scenarios. Left panel: Abundances of $\chi_1$, $\chi_2$ and $\chi_3$ in green, yellow and blue respectively in the C$_\chi 1_u$ scenario. The blue dashed line shows the equilibrium abundance of $\chi_3$. Right panel: Abundance of $\chi_3$ in blue and the sum of the abundances of $\chi_1$, $\chi_2$ and $\phi$ in green in the C$_\phi 1_u$ scenario. The dashed lines represent the equilibrium abundances.}
 \label{fig:abundance}
\end{figure}

In this appendix, we show our solutions of the Boltzmann equations from Eqs.~\eqref{eq:BME_full_chi} and \eqref{eq:BME_full_phi}, for the particle abundances as a function of the temperature parameter $x$ for two exemplary parameter points. The left and right panels of Fig.~\ref{fig:abundance} each show a representative point within the sub-scenarios C$_\chi 1_u$ and C$_\phi 1_u$, respectively. The former point is characterised by the masses $m_{\chi_1}=285.3\,\mathrm{GeV}$, $m_{\chi_2}=279.8\,\mathrm{GeV}$, $m_{\chi_3}=278.8\,\mathrm{GeV}$, $m_{\phi}\simeq570\,\mathrm{GeV}$ and a DM coupling $|\tilde{\lambda}_{u3}|=4.1\times 10^{-5}$. It corresponds to the one shown in the left panel of Fig.~\ref{fig:rescale_chi_u1} after requiring $\Omega h^2=0.12$. Annihilation is driven by $\chi_1$ which has a sizeable coupling to the top, $|\tilde{\lambda}_{t1}|=0.8$. Around $x\sim 100$, $\chi_1$ starts to decay into $\chi_2$ and $\chi_3$. Note that $\chi_2$ is rendered stable on the displayed time-scales -- due to the small mass splitting to $\chi_3$, $\Delta m_{23} = 0.38 \, \%$, in conjunction with the small DM coupling entering this decay rate -- but eventually decays into DM\@.

The chosen point for C$_\phi 1_u$ shown in the right panel of Fig.~\ref{fig:abundance} has a DM mass of  $m_{\chi_3}=451.1\,\mathrm{GeV}$ and mediator mass of $m_{\phi}=462.0\,\mathrm{GeV}$ while the states $\chi_1$ and $\chi_2$ are much heavier than $\phi$. Due to the sizeable couplings of the latter, their inter-conversions are fully efficient allowing us to consider the sum of their abundances in order to simplify the Boltzmann equations as mentioned in Sec.~\ref{sec:BME}. The DM abundance starts to depart from its equilibrium value already around $x\sim 3$ leading to a prolonged freeze-out process which is finally terminated by the decays of the remaining mediator particles into DM at $x\gtrsim 100$. The freeze-out dynamics are very similar to the one found in Ref.~\cite{Garny:2017rxs} considering a single fermionic state~$\chi$.

\bibliography{ref}
\bibliographystyle{JHEP}

\end{document}